\documentclass[journal]{new-aiaa} 
\usepackage[utf8]{inputenc}
\usepackage{lineno,hyperref,graphicx,subcaption}
\usepackage{graphicx}
\usepackage{amsmath}
\usepackage[version=4]{mhchem}
\usepackage{siunitx}
\usepackage{longtable,tabularx}
\usepackage{subcaption}
\usepackage[parfill]{parskip}
\graphicspath{ {./figures/} }

\title{Thrust Enhancement and Degradation Mechanisms due to Self-Induced Vibrations in Bio-inspired Flying Robots}
\author{Dipan Deb$^{*}$}
\author{Kevin Huang$^{+}$}
\author{Aakash Verma$^{o}$}
\author{Moatasem Fouda$^{o}$}
\author{Haithem E. Taha$^{\Diamond}$}

\affil{University of California Irvine, Irvine, California-92697, USA}

\begin{document}

\maketitle

\begin{abstract}
Bio-inspired flying robot (BIFR) which flies by flapping their wings experience continuously oscillating aerodynamic forces. These oscillations in the driving force cause vibrations in the motion of the body around the mean trajectory. That is, a BIFR hovering in place is not exactly stationary in space, rather it is oscillating in almost all directions about the hovering point in space. These oscillations affect the aerodynamic performance of the flier. Assessing the effect of these oscillations on particularly thrust generation in two-wings and four-wings BIFRs is the main objective of this work. To achieve such a goal two experimental setups were considered to measure the average thrust for the two BIFRs. The average thrust is measured over the flapping cycle of the BIFR. In the first experimental setup, the BIFR is installed at the end of a pendulum rod, in place of the pendulum mass. While flapping, the model creates a thrust force which raises the model along the circular trajectory of the pendulum mass to a certain angular position, which is an equilibrium point and it is also stable. Measuring the weight of the BIFR and the equilibrium angle it obtains, it is straightforward to estimate the average thrust by moment balance about the pendulum hinge. This pendulum setup allows the BIFR model to freely oscillate back and forth along the circular trajectory about the equilibrium position. As such, the estimated average thrust includes the effects of these self-induced vibrations. In contrast, we use another setup that relies on a load cell to measure thrust where the model is completely fixed.
The thrust measurement of both the BIFRs using the above mentioned setups exhibited an interesting trend. The load cell or the fixed test lead to a higher thrust than the pendulum or the oscillatory test for the two-wings model, showing an opposite behavior for the four-wings model. That is self induced vibration have different effects on the two BIFR models. Note that the aerodynamic mechanisms of thrust generation for two-wings and four-wings are fundamentally different. A two-wings BIFR generates thrust through traditional flapping mechanisms whereas a four-wings model enjoys clapping effect and wing-wing interaction. In the present work, we use motion capture system, aerodynamic modeling and flow visualization to study the underlying physics of the observed different behaviors of the two flapping models. The study revealed that the interaction of the vortices with the flapping wing robots majorly contributes to the observed aerodynamic behavior of the two BIFRs.

\end{abstract}

\section{Introduction}
Bio-inspired flying robots (BIFRs), more specifically Flapping Wing Micro Air Vehicle (FWMAV), have been a major focal point of research in the aerodynamics, dynamics, and control community in the last few decades In the twentieth century, the main attention was directed towards uncovering the unconventional lift mechanisms in flapping flight. With the more precise observation of the insect flight and how they make use of the unsteady lifting mechanisms (e.g., wake capture, leading edge vortex, etc.), this puzzle was resolved \cite{LEVinsects} \cite{dickinson1999wing} \cite{Ellington1984}. Having understood the lifting mechanisms in insect flight, several researchers independently developed unique designs for FWMAVs \cite{DARPA1NAV} \cite{hassanalian2017classifications}. Zakaria et al. (2015) showed that the inclusion of intertial power requirements is essential for physical and proper optimization \cite{Pterosaur}. They further studied the aerodynamic forces generated in forward flight for different Reynolds numbers and flapping frequency \cite{Zakaria}. Whitney (2012)  designed and developed one of the tiniest FWMAVs \cite{whitney2012design}. Keennon et al. (2012) developed the aerovironment nano humming-bird as an ornithopter that hovers \cite{keennon2012development}. TU Delft researchers designed and developed, studied aerodynamic and dynamic performances of the Delfly \cite{de2012delflydesign} \cite{de2016delfly}. Any design like the previously mentioned ones is aimed at a particular objective or flying condition like forward flight, hovering etc. 

One of the most challenging flying conditions of flapping flight is hovering. There have been numerous studies to investigate the aerodynamics of hovering flapping flight. Weis-Fogh tested the quasi-steady assumption for insect flight where unsteady effects are more conspicuous and showed that quasi-steady aerodynamics can predict the main features of hovering flights \cite{Weisfogh1972}. However, Ellington \cite{Ellington1} examined the results of Weis-Fogh's in the light of more accurate kinematic and morphological data, and his conclusion was opposite to that of Weis-Fogh's. Ellignton \cite{Ellignton4} further showed that leading edge separation bubble plays a prominent role in the hovering flight of insects. In comparison to a thin airfoil, Ellington asserted that the leading edge bubble modifies the camber and the thickness of the thin profile which enhances lift at low Reynolds numbers. Bayiz et al. (2018) \cite{Bayiz_2018} compared the hovering efficiency in rotary and flapping modes using rigid rectangular wings. They observed that flapping wings are more efficient in achieving higher average lift coefficient in hovering. Sarkar et al. \cite{SARKAR201372} studied aerodynamic performance under asymmetric flapping kinematics using both sinusoidal and triangular waveforms. The frequency-asymmetry mechanism showed an increase in aerodynamic loads for the sinusoidal case. During the faster stroke, lift can be enhanced depending on the level of asymmetry. The results of these investigations can be used to design an efficient flapping robot for hovering. 

Discussion on hovering insects is incomplete without pondering upon the question of stability. Sun et al. \cite{sun2007dynamic} found that pitching moment produced by change in horizontal speed is the primary source of an unstable oscillatory mode, where as vertical force produced by changes in vertical speed is the primary source of a stable slow subsidence mode. These results are mainly based on averaging the flight dynamics over the flapping cycle. In contrast, using chronological calculus a hidden stabilization mechanism was discovered in a hovering hawkmoth. Taha et al. \cite{hiddenstability} showed that insects use a passive stabilization mechanism through their natural wing oscillation; this is called vibrational stabilization. It is a natural phenomenon observed in systems like Kapitza pendulum \cite{kapitza1951dynamic}. A bio-inspired flapping robot in a two degree of freedom system can exhibit vibrational stabilization as well \cite{kianistable}. 

It is important to emphasize that when an insect hovers over a flower, it is not completely stationary in space over the flower. Rather, it oscillates in all directions. This oscillation can be observed in the video Hedrick and Daniel (2006) shown in the supplementary section of their paper \cite{hawkmoth}. Hence it experiences self-induced vibrations. This vibration may change the flow field around the wing and thus affect the generated aerodynamic forces. In ideal hovering, there should be no self-induced vibration but in real cases these vibrations are unavoidable due to the inevitable oscillatory nature of the driving aerodynamic forces. To the authors' best knowledge, there is little effort exerted that focuses on this point \cite{higher_order_2015} \cite{hoveringofmodelinsects}, and almost no effort exerted to study this effect on clapping effects. The current work is dedicated to study the effect of  self-induced body vibration in flapping flight. To achieve this goal, we considered two different setup for aerodynamic force measurement (specially thrust) : (1) Pendulum setup or oscillatory test and (2) Load cell setup or fixed test. In the pendulum setup, we replaced the mass of a pendulum with a flapping wing robot. Whenever the robot flaps, it generates thrust and moves upward along the circular trajectory of the pendulum, assuming an equilibrium at some angular position. We can measure this angle and use it to calculate the averaged thrust. The robot vibrates about this angular position i.e., the measured thrust includes the effect of vibration. In contrast, in the loadcell or fixed test setup, there is no room for such a vibration. So using these two setups, we can measure the effect of self-induced vibration on flapping thrust generation. 

Two flapping wing robots are considered for this study: one has two wings and the other has four wings. One of them is a two-wings model and it is also named Model A (Figures \ref{fig:ModAtop} \ref{fig:ModAside} \ref{fig:ModAcorner}  \ref{fig:realmechAside}). The wings of these models have stroke angle of $\sim 60^{0}$. Unsteady responses like leading edge vortex \cite{LEVinsects}, wake capture \cite{dickinson1999wing} \cite{sane2003} are utilized by the two-wings model in flapping flight. The four-wings model or Model B (Figures \ref{fig:ModBtop}  \ref{fig:ModBside} \ref{fig:ModBcorner}) exploits an wing-wing interaction phenomenon named 'clap-and-peel' for generating thrust. There has been a surge of interest in the recent years to study the interaction of multiple bodies in a fluid flow  \cite{Yaowei2020} \cite{DEB2020103133} \cite{KIM201918}. Deb et al.(2020) \cite{DEB2020103133} observed in-phase and out-of-phase oscillations in a couple of rigid plates which are oriented one beside another, through wind tunnel experiments. They explained the modes of oscillation of the plates through flow field data and visualization. Some studies also included flexibility. Flexible flags in different orientations like tandem and side-by-side and their interaction in those configurations were studied by Alben (2009) \cite{alben_2009}. In flapping flight, wing-wing interaction has been exploited by the four-wings model by using 'clap-and-fling' \cite{jadhav2019effect} (for flexible wings -'clap-and-peel' \cite{armanini2017}) to generate thrust. Outcomes of wing-wing interaction like stronger leading edge vortex \cite{armanini2016quasi} and jet effect \cite{jadhav2019effect} \cite{sane2005wingwing} are utilized for thrust generation by the clap-and-peel mechanism. Balta et al. \cite{MiquelDipan2021} showed with flow visualization that the peel phase of the flapping cycle draws air in and the clap phase propels it downstream; that thrust is augmented using jet effect. A blob of air flows between the wings in the peel phase, which strengthens the leading edge vortex. Some efforts were made to capture this effect into an aerodynamic model \cite{lighthill_1973} \cite{kolomenskiy_moffatt_farge_schneider_2011} \cite{WU1984}. Armanini et al. \cite{armanini2016quasi} studied the improved strength of the leading edge vortex and included clapping effect into a quasi-steady aerodynamic model. So the mechanism for thrust generation in a four-wings robot is fundamentally different from the conventional one due to the wing-wing interaction, or the clap-and-peel mechanism. Thus it is expected that the effects of the self induced vibration will not be similar for the two-wings and the four-wings.

In the present work, we investigated the effect of self-induced body vibration on two-wings and four-wings (clap-and-peel) flapping robots using two different thrust measurement setups, flow visualization, motion capture and theoretical aerodynamic modeling.

\section{Observation and Experimental Setup}

\begin{figure}
	\begin{subfigure}{.4\textwidth}
		\centering
		\includegraphics[width=1\linewidth]{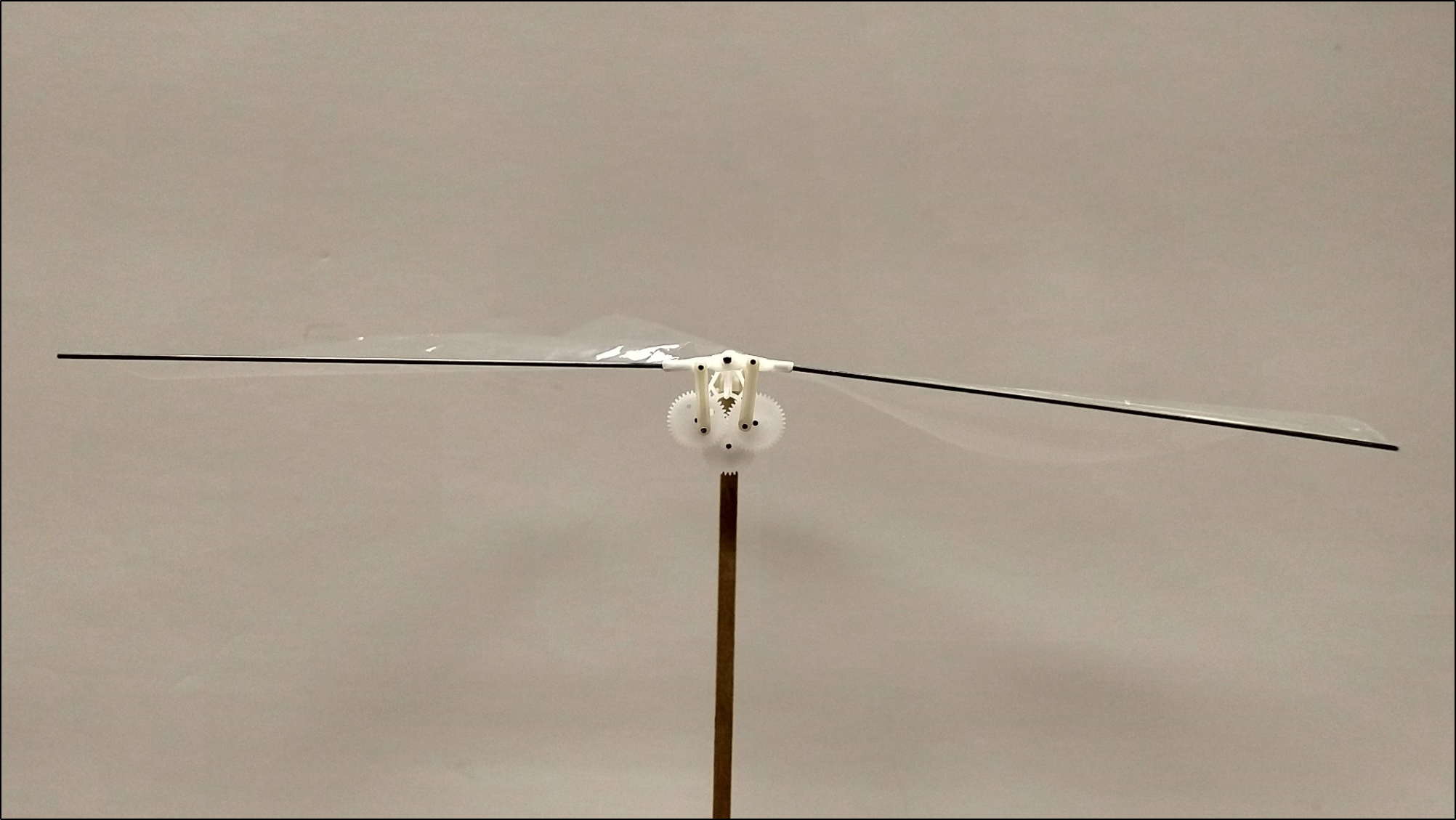}
		\caption{Model A}
		\label{fig:ModAtop}
	\end{subfigure}%
	\hfill	
	\begin{subfigure}{.4\textwidth}
	 	\centering
		\includegraphics[width=1\linewidth]{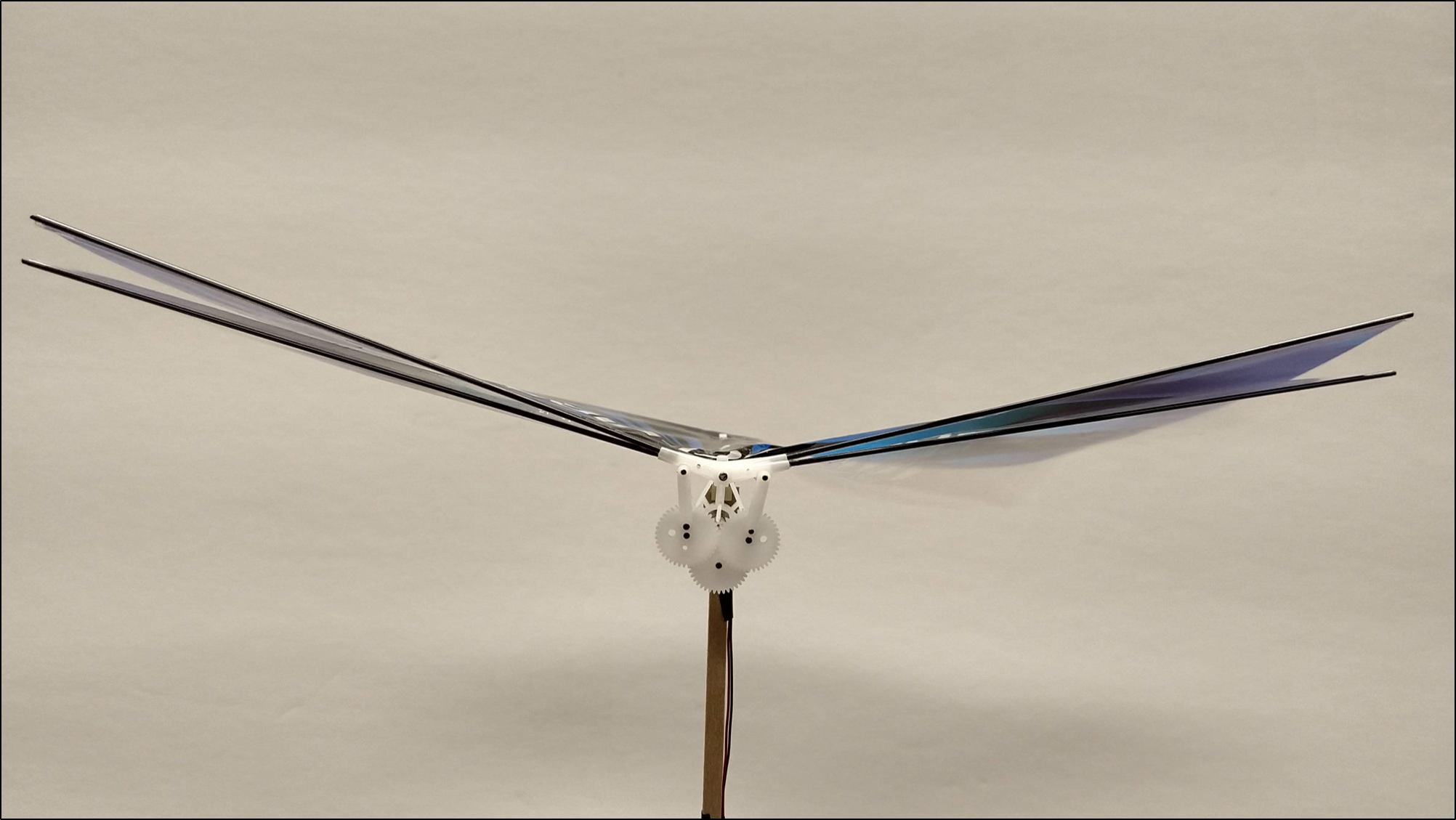}
		\caption{Model B}
		\label{fig:ModBtop}
	\end{subfigure}%	
	\caption{View from the front of the robots which were also used by Balta et al. (2021) \cite{MiquelDipan2021}}
	\label{fig:modelstop}
\end{figure}

\begin{figure}
	\begin{subfigure}{.4\textwidth}
		\centering
		\includegraphics[width=1\linewidth]{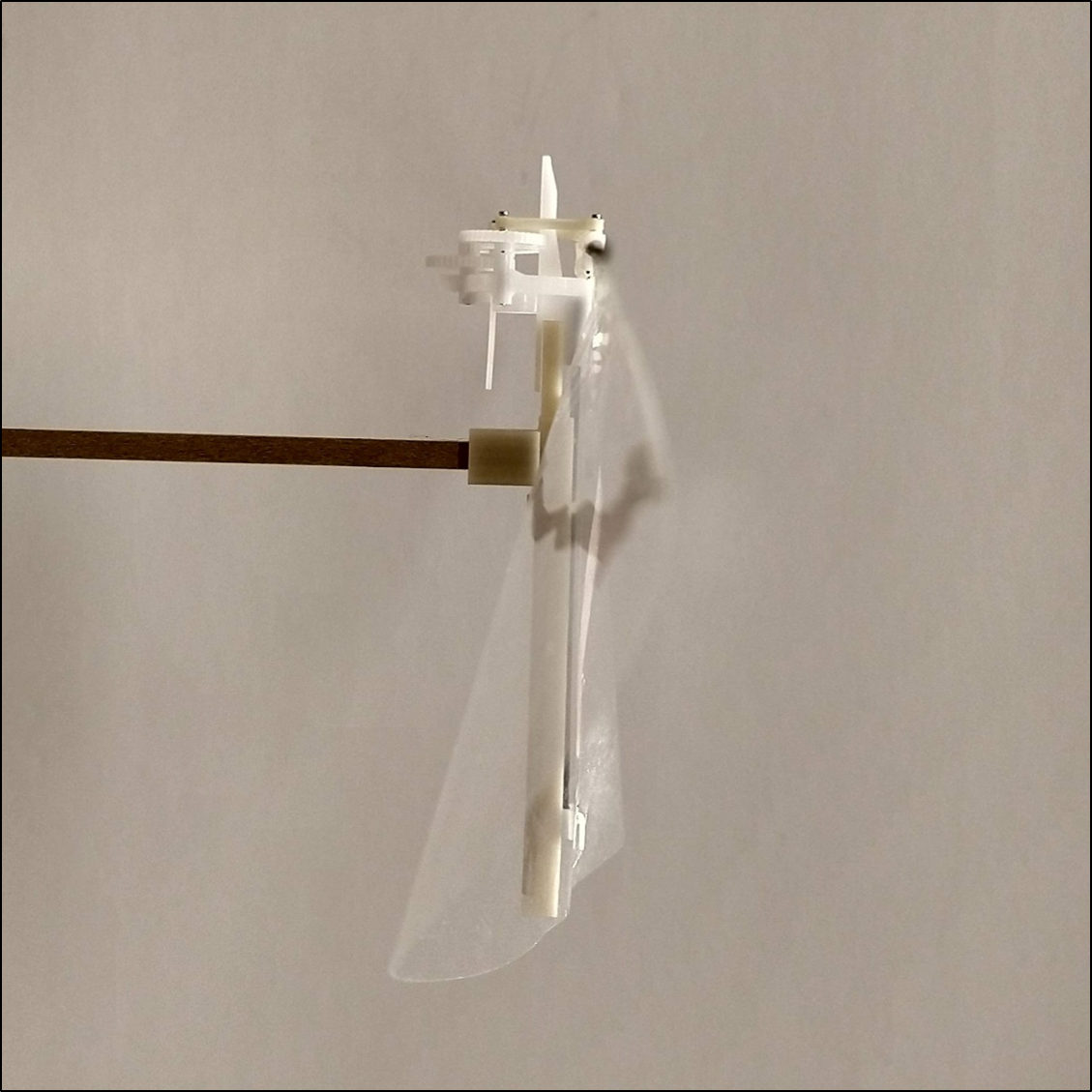}
		\caption{Model A}
		\label{fig:ModAside}
	\end{subfigure}%
	\hfill	
	\begin{subfigure}{.4\textwidth}
	 	\centering
		\includegraphics[width=1\linewidth]{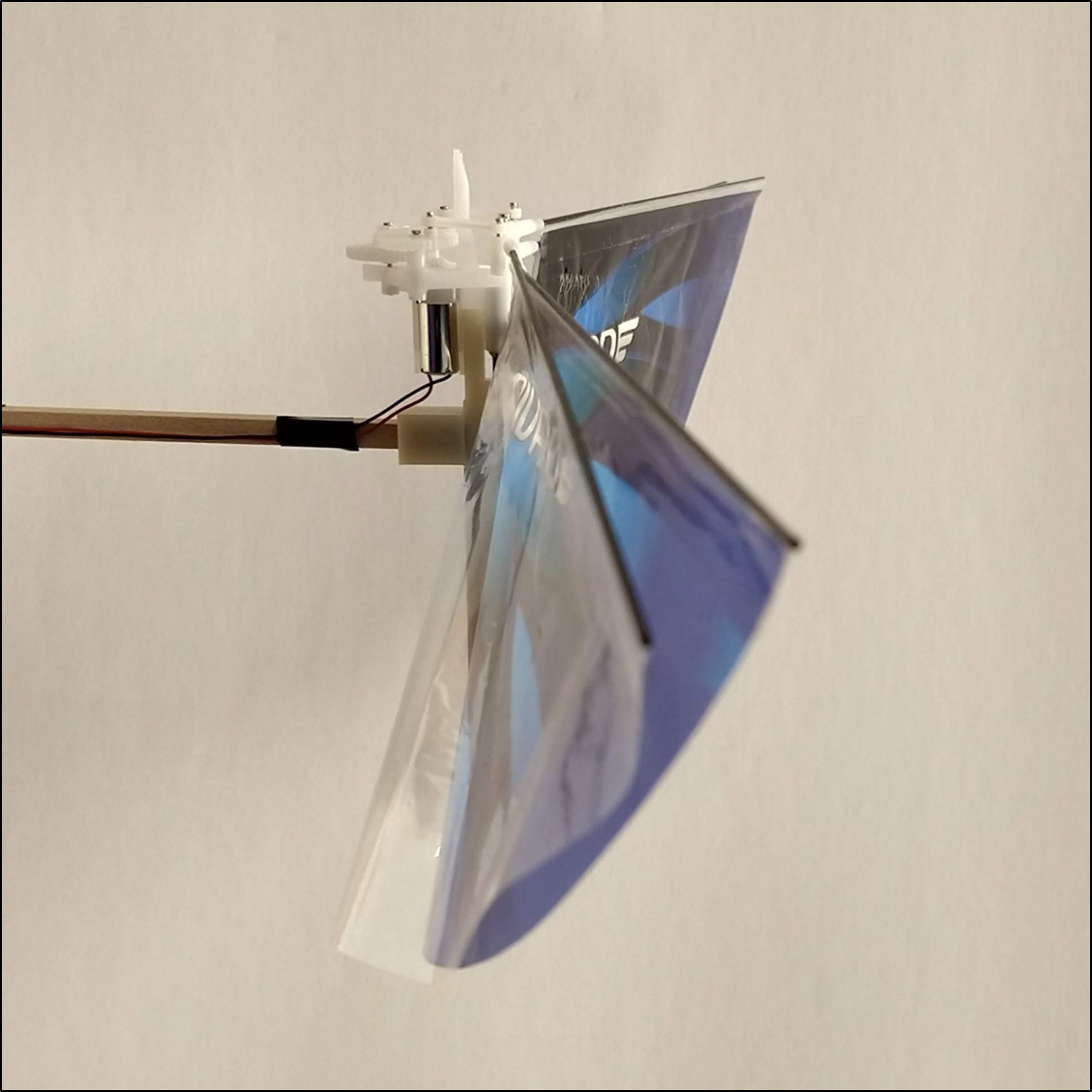}
		\caption{Model B}
		\label{fig:ModBside}
	\end{subfigure}%	
	\caption{View from the side of the FWMAVs \cite{MiquelDipan2021}}
	\label{fig:modelsside}
\end{figure}

\begin{figure}
	\begin{subfigure}{.4\textwidth}
		\centering
		\includegraphics[width=1\linewidth]{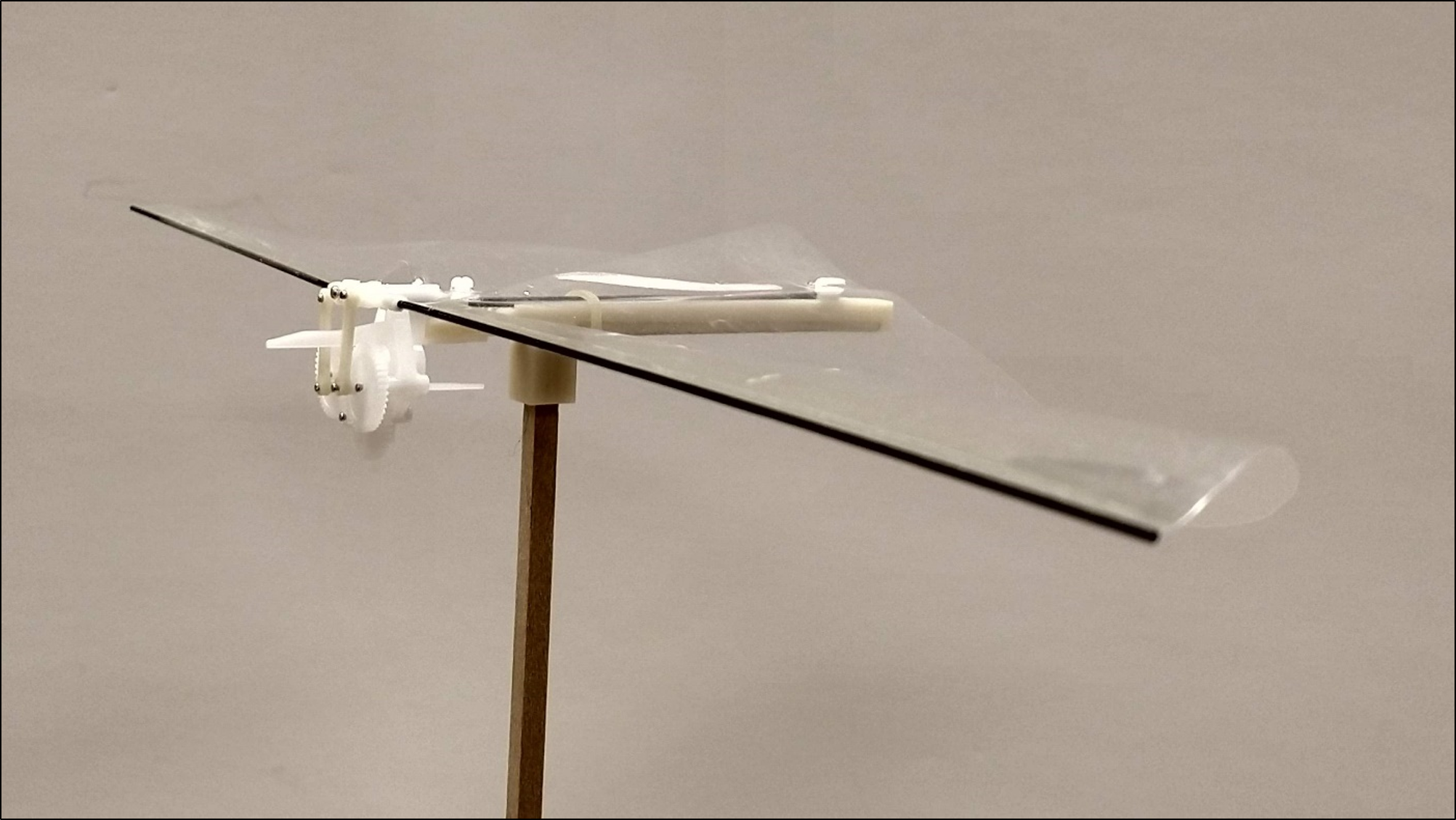}
		\caption{Model A}
		\label{fig:ModAcorner}
	\end{subfigure}%
	\hfill	
	\begin{subfigure}{.4\textwidth}
	 	\centering
		\includegraphics[width=1\linewidth]{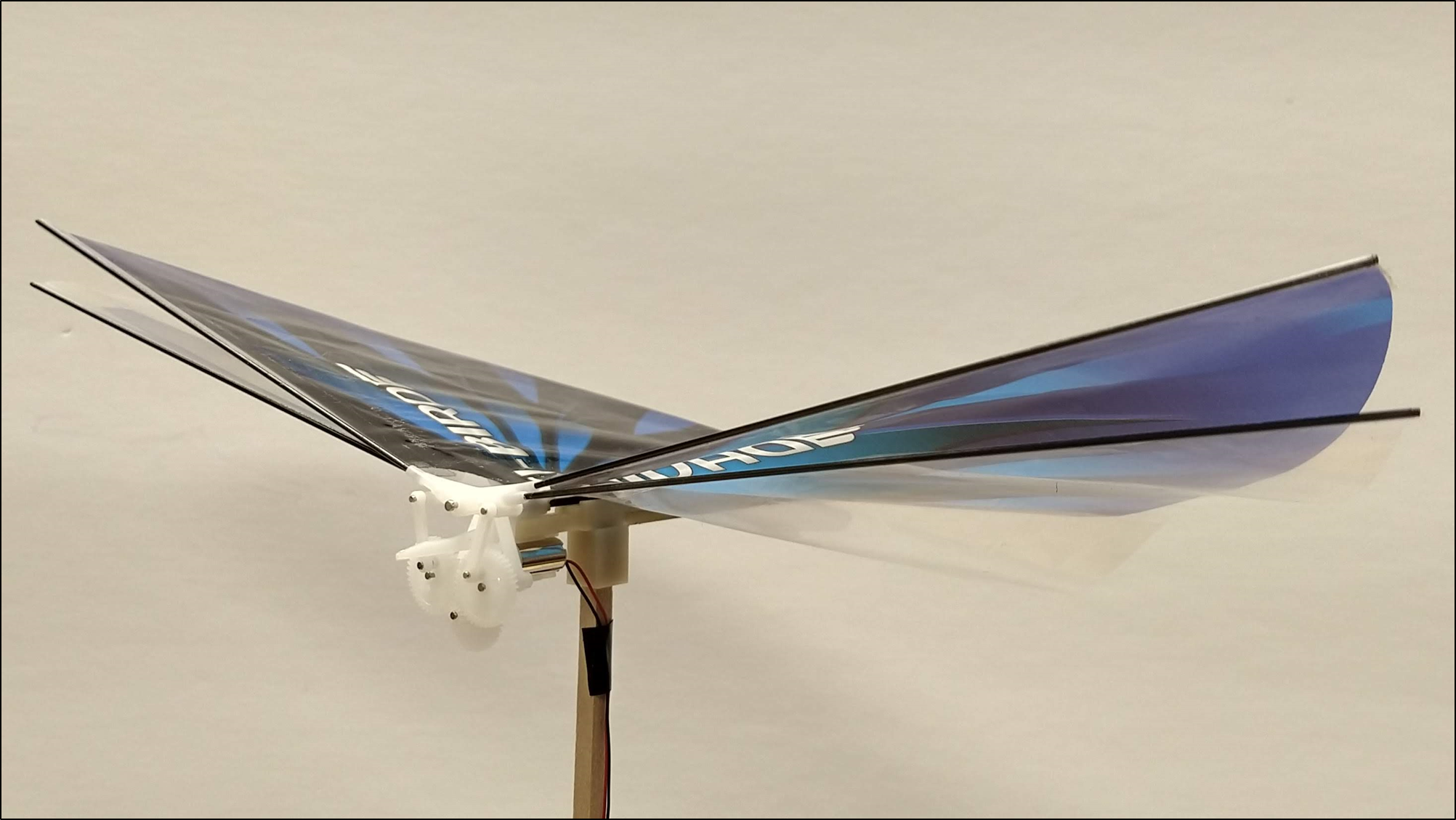}
		\caption{Model B}
		\label{fig:ModBcorner}
	\end{subfigure}%	
	\caption{View from the corner of the FWMAVs \cite{MiquelDipan2021}}
	\label{fig:modelscorner}
\end{figure}

\begin{figure}
		\centering
		\includegraphics[width=0.6\linewidth]{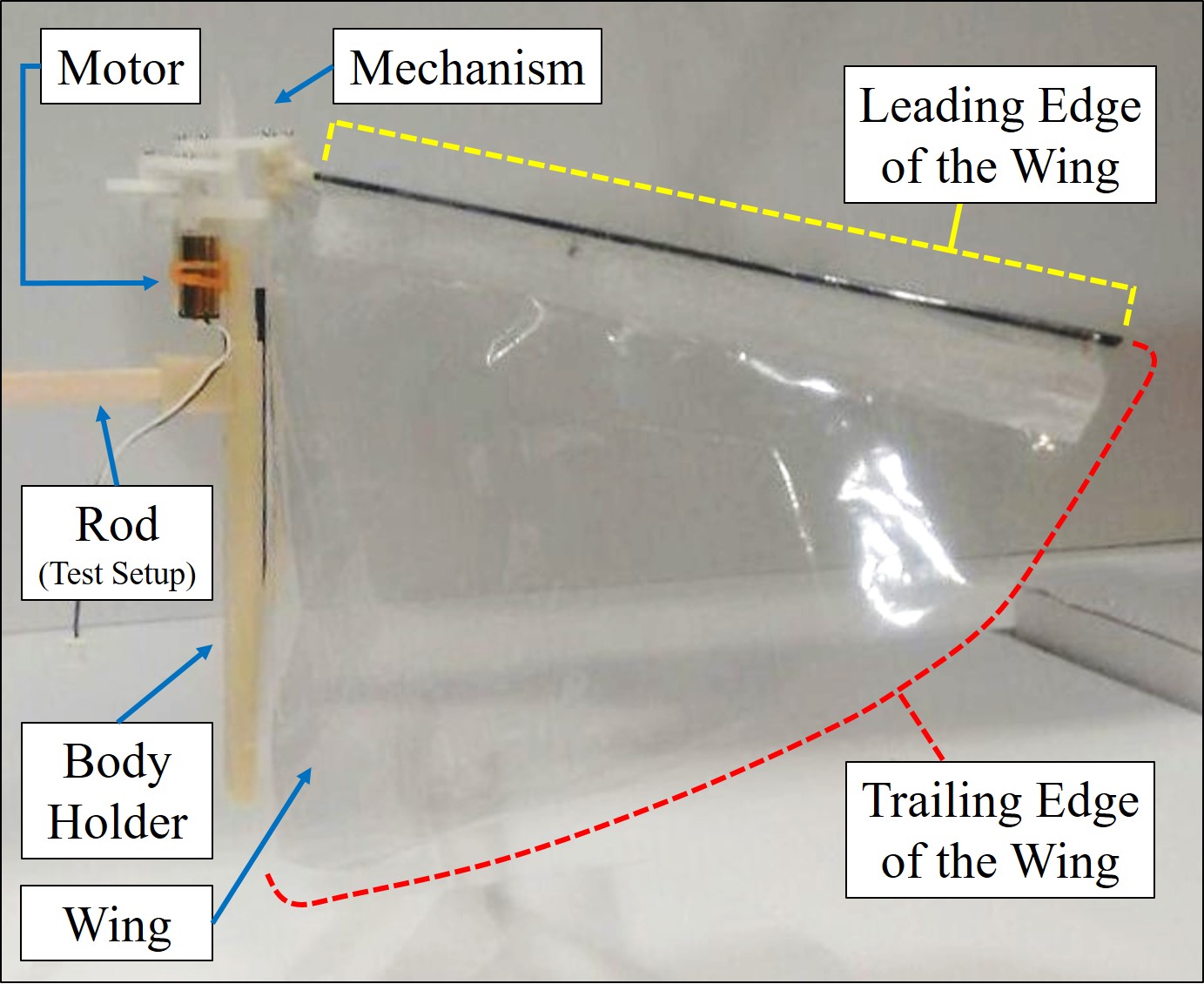}
		\caption{Detailing and different components of Model A \cite{MiquelDipan2021}}
		\label{fig:realmechAside}
\end{figure}

This focus of this research is to investigate the effect of self-induced oscillations on two distinct mechanisms of thrust generation for FWMAVs. The flapping robots, Model A (Figures \ref{fig:ModAtop}, \ref{fig:ModAside}, \ref{fig:ModAcorner}) and Model B (Figures  \ref{fig:ModBtop}, \ref{fig:ModBside}, \ref{fig:ModBcorner}), utilize a crank-rocker mechanism for flapping, which is also used in the same laboratory by Balta et al. (2021) \cite{MiquelDipan2021}. In case of Model B, wing-wing interaction takes place which exploits the clap-and-peel mechanism. The motion of the wings moving towards each other during the flapping cycle, is called clap; and the ensuing motion of moving away from each other is known as peel. Given the flexibility of the wings, they peel away in this motion and hence the name. Figure \ref{fig:realmechAside} shows leading edge (LE), trailing edge (TE), motor and other parts of the mechanism.

% Pictures of the real Models from the top

% Pictures of the real Models from the side

% Pictures of the real Models from a corner

\begin{figure}
	\begin{subfigure}{.45\textwidth}
		\centering
		\includegraphics[width=1.2\linewidth]{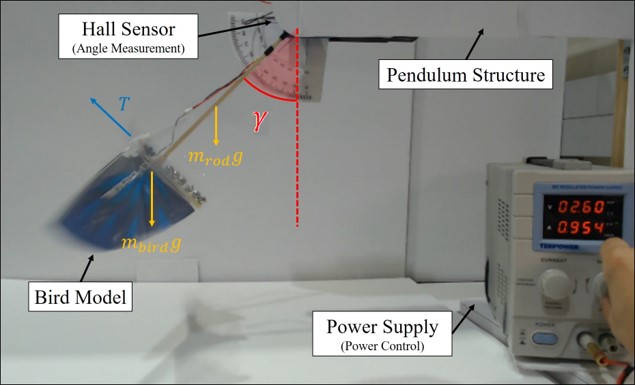}
		\caption{Pendulum test which was also used by Balta et al.(2021) \cite{MiquelDipan2021}}
		\label{fig:pn}
	\end{subfigure}%
	\hfill	
	\begin{subfigure}{.45\textwidth}
	 	\centering
		\includegraphics[width=0.8\linewidth]{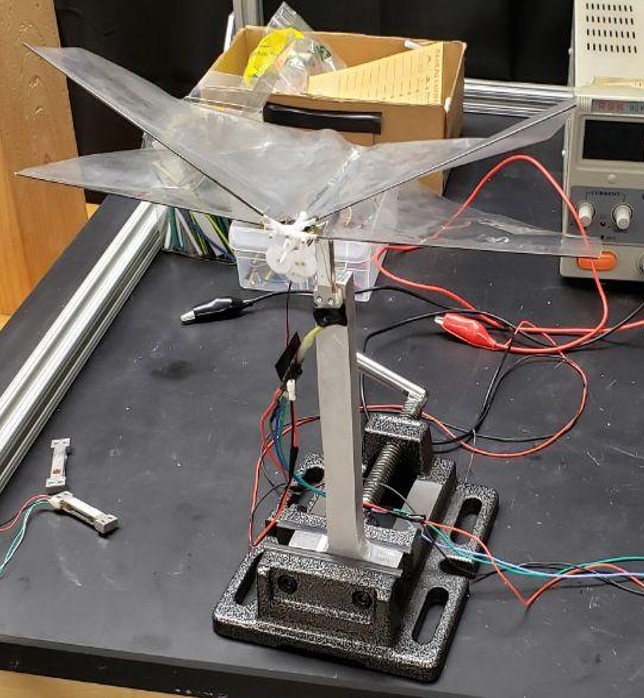}
		\caption{FWMAV mounted on a loadcell setup}
		\label{fig:lc}
	\end{subfigure}%	
	\caption{The different setups for aerodynamic force measurements}
	\label{fig:setups}
\end{figure}

The present work is dedicated to testing and analyzing Models A \& B. We use different experimental setups like (1) Pendulum test or oscillatory test (Figure \ref{fig:pn}) and (2) Loadcell or fixed test (Figure \ref{fig:lc}) for measuring the thrust generation by the two above mentioned models. We trim an aluminium block to fit the loadcell onto it and we mount the FWMAV on the loadcell. The thrust signals from the loadcell is filtered and acquired through a low pass filter and NI DAQ respectively.  

The pendulum experiment consists of an encoder. The model is connected to a rigid wooden rod. When the power supply is turned off or the no power is applied to the model, it can relax and does assume a resting state. When the FWMAV attains a certain flapping frequency, it generates thrust and moves upward along the circular arc of the pendulum, as shown in Figure \ref{fig:pn}. So, it attains a stable equilibrium point at some angle $\gamma$. The pendulum angle $\gamma$ can be measured using the encoder. By measuring the mass of the wooden pendulum rod (denoted by $m_{r}$) and of the FWMAV (denoted by $m$) we can calculate the average thrust over the flapping cycle. We obtained the equation \ref{eqn:Thrust} for average thrust calculation with moment balance about the pendulum hinge.

\begin{equation}\label{eqn:Thrust}
T = \left( m+\frac{1}{2}m_{\rm{r}} \right) g \sin\gamma,
\end{equation}

Using this method and the equation we can measure mean thrust for a given value of the flapping frequency. A stroboscope was used to measure this frequency. In this test the FWMAV body is free to oscillate. So, the effect of body-induced vibrations are already included in the measured average thrust.

On the other hand, the load cell test (Figure \ref{fig:lc}) measures thrust with time during the flapping cycle. By applying time average on the obtained data we can calculate mean thrust at each given value of the flapping frequency. The loadcell setup consists of a uni-axial loadcell. We use LabVIEW for obtaining and processing the loadcell data. 

\begin{figure}
	\begin{subfigure}[b]{.5\linewidth}
		\centering
		\includegraphics[width=.8\linewidth]{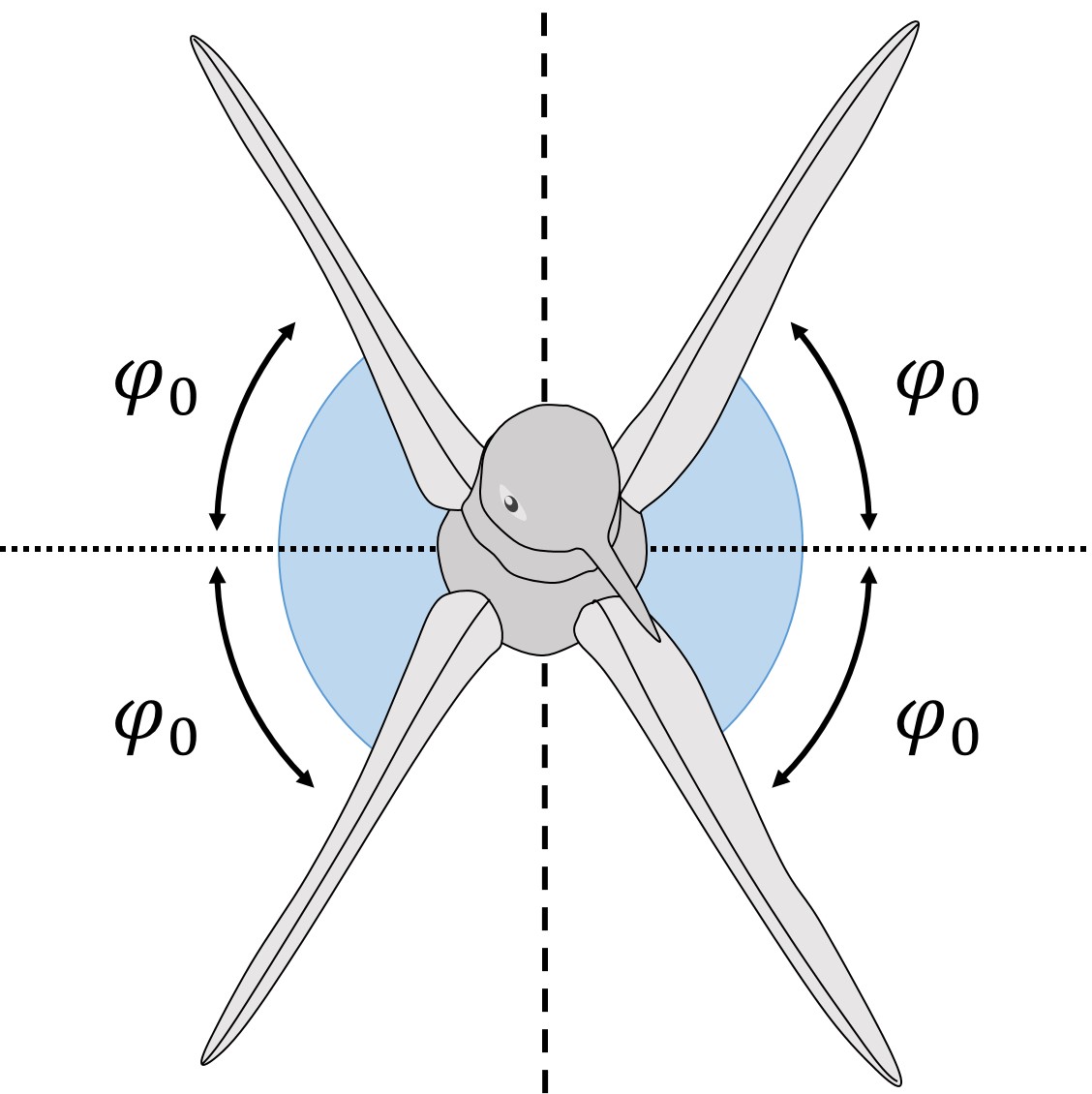}
		\caption{Flapping robot with four wings (Model B)}
		\label{fig:bird_4_stroke}
	\end{subfigure}%
	\hfill
	\begin{subfigure}[b]{.5\textwidth}
		\centering
		\includegraphics[width=.8\linewidth]{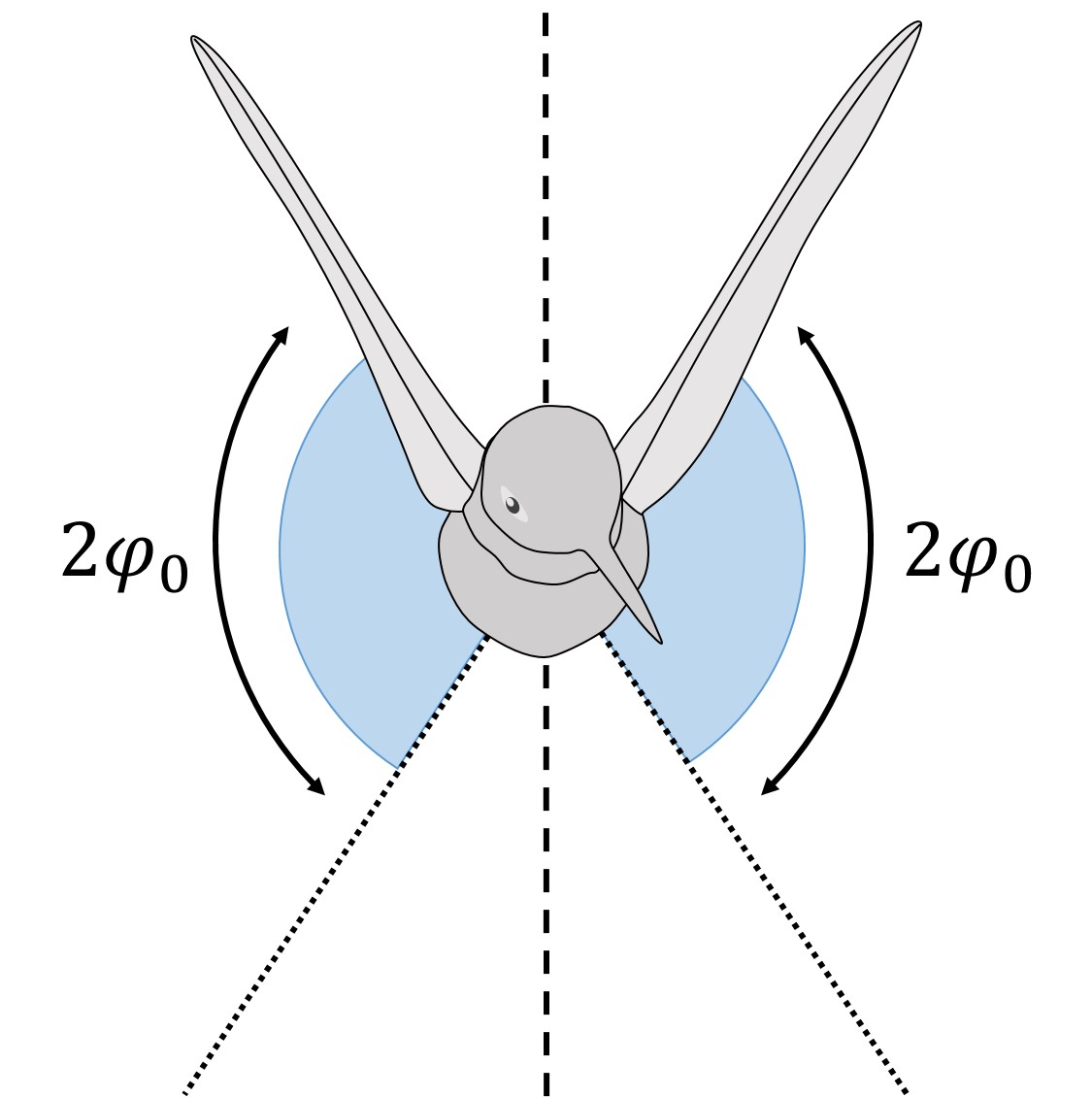}
		\caption{Flapping robot with two wings (Model A)}
		\label{fig:bird_2_stroke}
	\end{subfigure}%
	
	\caption{Amplitude of flapping angle of an individual wing corresponding to the flapping robot \cite{MiquelDipan2021}}
	\label{fig:strokewings}
\end{figure}

We non-dimensionalized the measured thrust for dynamic similarity. However, it is important to note that the generated thrust depends on the angle swept by the wings, wing surface area, flapping frequency, number of wings, and wing-span. So, we non-dimensionalize the thrust force by $\frac{1}{2}\rho V_{ref}^2 S N$, where $V_{ref}=2\pi f R \Phi$ is a reference speed, taken here the maximum speed of the wing tip, similar to helicopters and propellers. Also $f$ denotes flapping frequency, $R$ denotes wing span and $\Phi$ is amplitude of flapping angle for a single wing. Figure \ref{fig:bird_4_stroke} shows that angle swept by one wing for Model B is almost half as much for its Model A equivalence. 
So, $\Phi=2\phi_{0} \& N=2$ are for Model A and similarly we can say that $\Phi=\phi_{0} \& N=4$ are for Model B.
For Model A the coefficient of thrust is defined as
\begin{equation}\label{th_2wings}
    C_{T}=\frac{T}{\frac{1}{2}\rho (2\pi f R 2\phi_{0})^2 2 S }
\end{equation}

whereas for Model B it can written as,
\begin{equation}\label{th_4wings}
    C_{T}=\frac{T}{\frac{1}{2}\rho (2\pi f R \phi_{0})^2 4 S }
\end{equation}

\begin{figure}
	\begin{subfigure}[b]{.5\linewidth}
		\centering
		\includegraphics[width=1\linewidth]{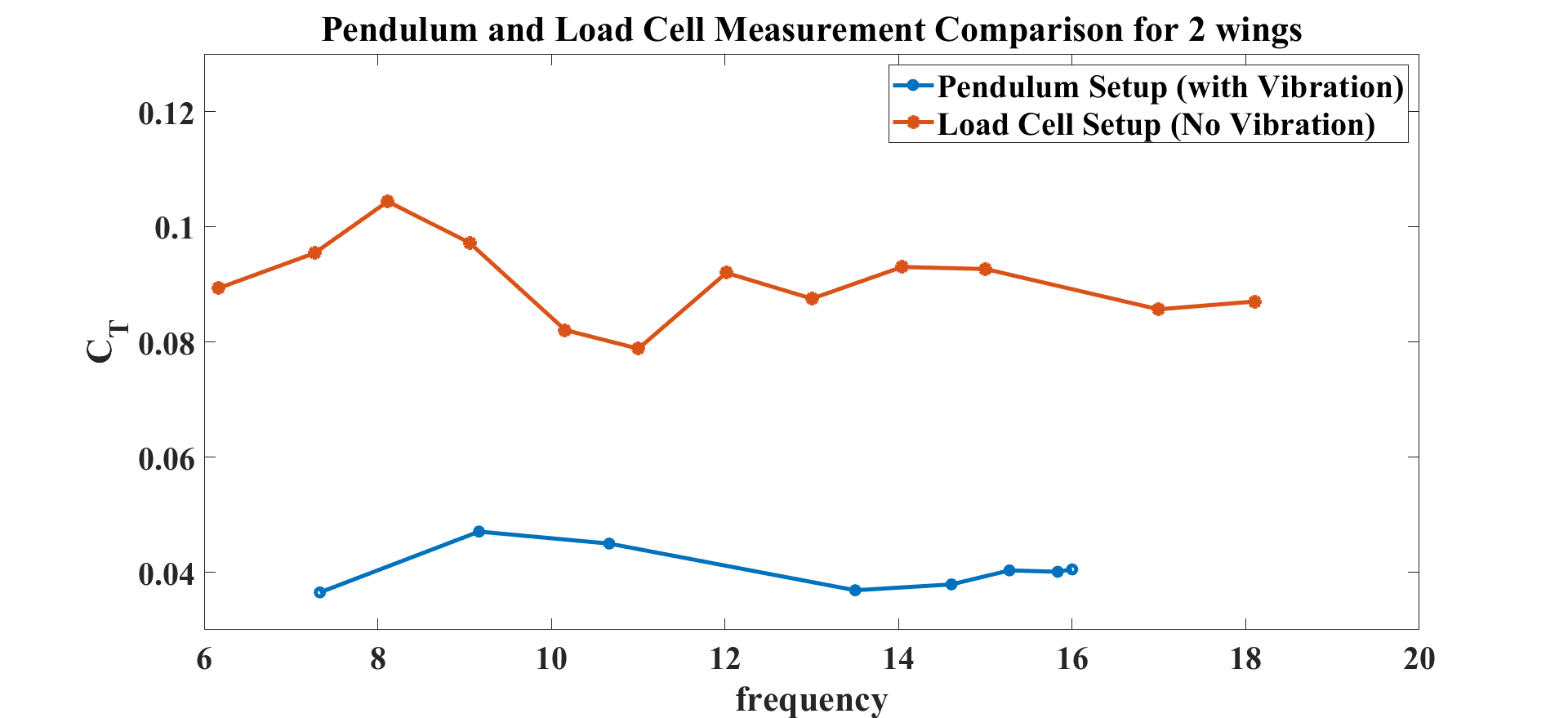}
		\caption{$C_{T}$ vs f for Model A}
		\label{fig:2T}
	\end{subfigure}%
	\hfill
	\begin{subfigure}[b]{.5\textwidth}
		\centering
		\includegraphics[width=1\linewidth]{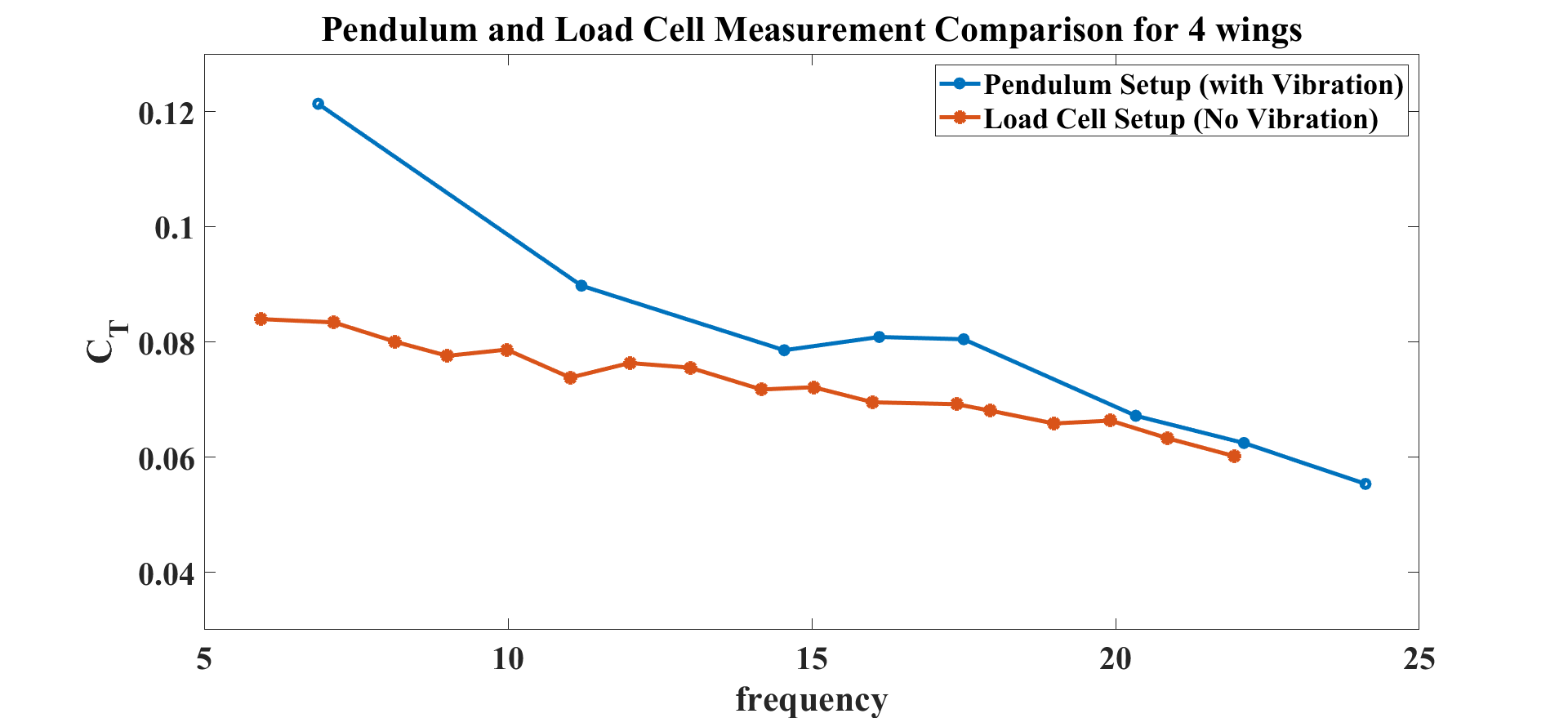}
		\caption{$C_{T}$ vs f for Model B}
		\label{fig:4T}
	\end{subfigure}%
	
	\caption{Comparison of thrust co-efficient for both the Models (A \& B) from both the pendulum and loadcell setups}
	\label{fig:observation}
\end{figure}

Figure \ref{fig:2T} shows the averaged thrust coefficient $C_{T}$ measured for model A for given flapping frequencies using the experimental setups - fixed test or the test without any vibration and oscillatory test or the test with vibration. It clearly shows that the thrust measurements from the oscillatory test are less than the fixed test. On the other hand, the situation is reversed for Model B as shown in Figure \ref{fig:4T}. This clear difference in behaviour is the main focus of this paper. The generated aerodynamic forces by the FWMAVs are periodic in nature. Hence even after achieving a stable equilibrium in the system, the FWMAV oscillates about that equilibrium point. On the other end, the fixed test setup allows no such oscillation. Moreover, Model B exploits wing-wing interaction or clapping effect to generate thrust which differs from the thrust generation of Model A \cite{MiquelDipan2021}. Thus the vibration also has different effects on the thrust generation and results in the opposite trend observed in Figure \ref{fig:observation}.

\begin{figure}
		\centering
		\includegraphics[width=0.5\linewidth]{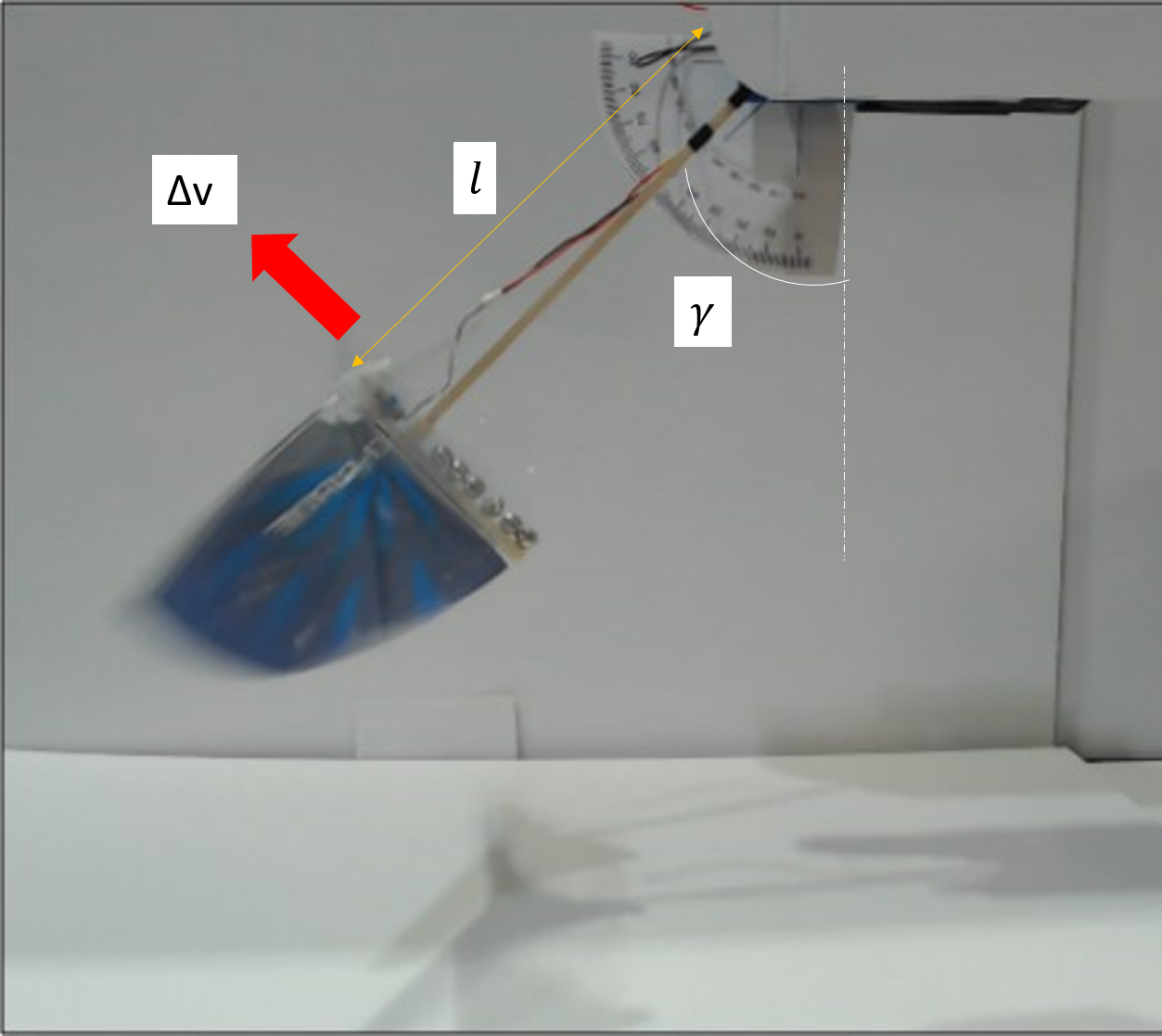}
		\caption{$\Delta v$ measurement from the Pendulum angle}
		\label{fig:delv1}
\end{figure}

\begin{figure}
	\begin{subfigure}[b]{.5\linewidth}
		\centering
		\includegraphics[width=0.7\linewidth]{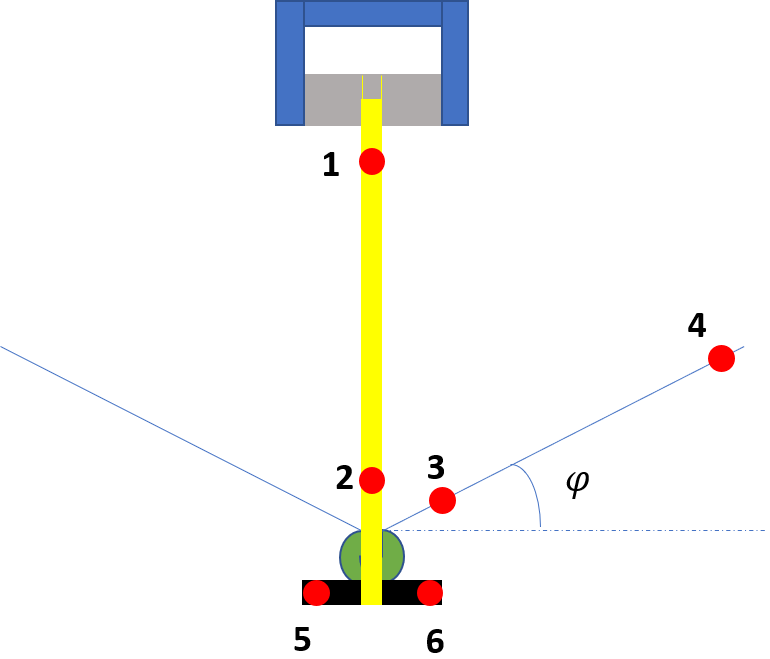}
		\caption{View from the front}
		\label{fig:frontview}
	\end{subfigure}%
	\hfill
	\begin{subfigure}[b]{.5\textwidth}
		\centering
		\includegraphics[width=0.7\linewidth]{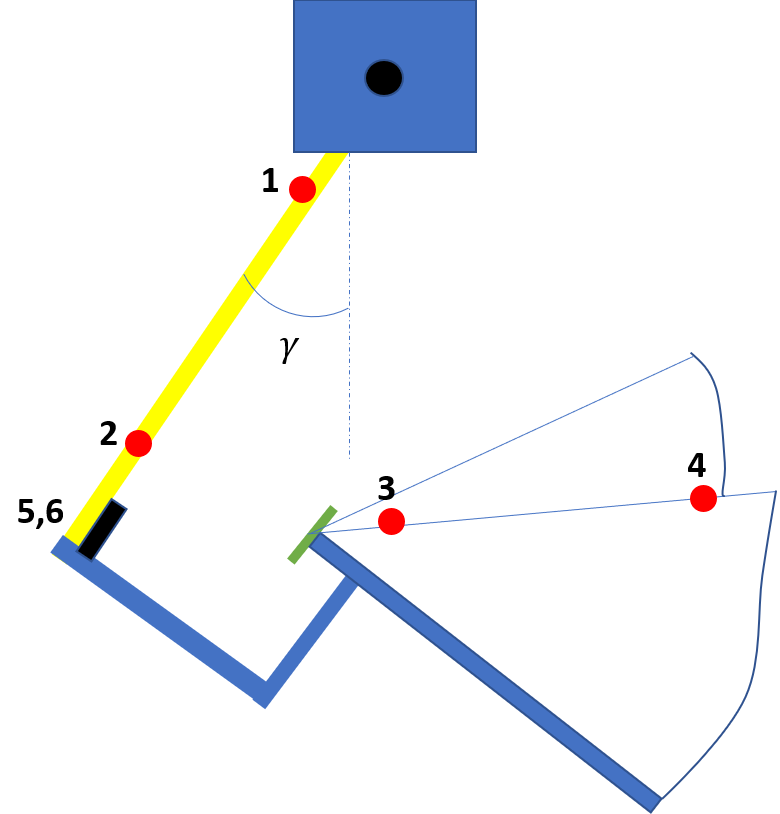}
		\caption{View from the side}
		\label{fig:sideview}
	\end{subfigure}%
	\caption{Schematic of the FWMAV and the pendulum with the active markers}
	\label{fig:markers}
\end{figure}

In order to investigate the effect of the vibration, we need to define it. In the oscillatory test, the angular position of the FWMAV, denoted by $\gamma$, oscillates around a mean point $\gamma_{0}$, with a zero-mean periodic variation $\widetilde{\gamma}$. For a given flapping frequency we can say that the FWMAV assumes an angular location $\gamma (t) = \gamma_{0}+\widetilde{\gamma}(t)$. Denoting the length of the wooden rod as $l$, the vibration velocity can be defined as $\Delta v= l \Dot{\gamma}$. The red arrow shows the direction of positive $\Delta v$ in the Figure \ref{fig:delv1}. The flapping angle $\phi$ is also measured simultaneously with the pendulum angle $\gamma$ at a given point in time. For this purpose, we used a motion capture system with one tracker and six markers as mentioned in this 2022 conference paper \cite{selfinduced}. Figure \ref{fig:markers} shows the positions of these markers. Figure \ref{fig:sideview} shows that the markers positioned on the wooden rod (1 \& 2) are for measuring $\gamma (t)$, the markers on the leading edge (3 \& 4) as shown in Figure \ref{fig:frontview} are for the measurement of $\phi (t)$, and the two markers (5 \& 6) at the bottom of the rod are for defining local horizon. All these markers were active in nature. A 3D tracker receives signals from the markers and sends them to the computer. The tracker has 0.1mm spatial and $1\mu s$ temporal. The VZSoft software acquires the signal and records the data on the computer. We used MATLAB to read and process the data to observe the perturbation motion.

\begin{figure}
	\begin{subfigure}[b]{.5\linewidth}
		\centering
		\includegraphics[width=0.9\linewidth]{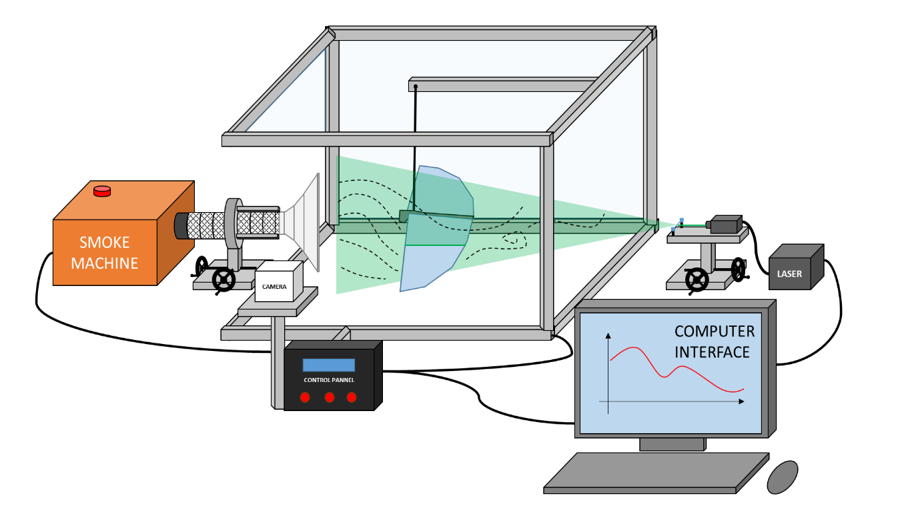}
		\caption{Schematic of the setup for Flow Visualization \cite{MiquelDipan2021}}
		\label{fig:flowviz}
	\end{subfigure}%
	\hfill
	\begin{subfigure}[b]{.5\textwidth}
		\centering
		\includegraphics[width=0.9\linewidth]{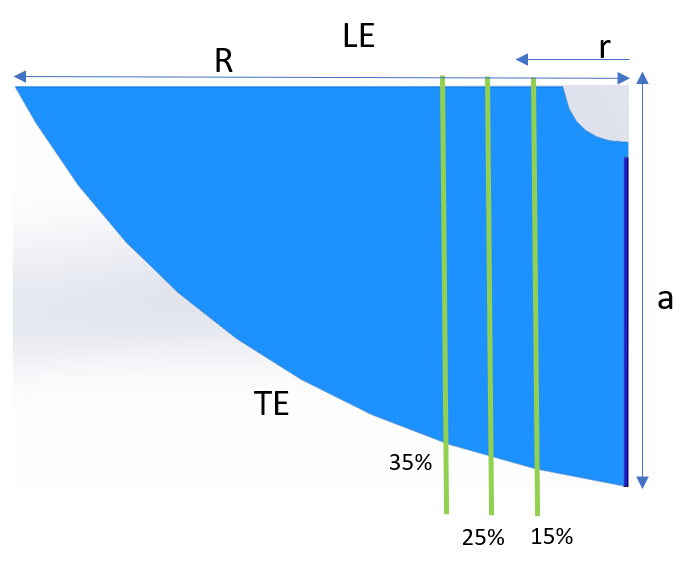}
		\caption{Visualization locations in the spnawise direction}
		\label{fig:section1}
	\end{subfigure}%
	
	\caption{Schematic of the Flow Visualization setup and the spanwise sections on the wing for visualization}
	\label{fig:Visualization}
\end{figure}

Flow visualization is executed to investigate the effect of the vibration in the flow field for both the models and setups. The images captured for the cases with and without vibration to explain the flow physics underlying the performance observed in Figure \ref{fig:observation}. Figure \ref{fig:Visualization} shows a schematic of the flow visualization setup. Figure \ref{fig:flowviz} shows the FWMAV being mounted near a diffuser, that is attached to a smoke machine. The diffuser is used to inject fog by the machine into the flow. The fog follows the flow field generated by the flapping of models. Specific planar sections of the flow field were enlightened by a laser sheet generated by a class III laser machine. The visualization was captured for 6 Hz flapping frequency with a camera at 240 FPS. As the laser sheet is two dimensional, the visualization is done at different spanwise positions. These locations are demarcated in green in Figure \ref{fig:section1}.

\newpage

\section{Aerodynamic Modeling}
To further understand the effect of perturbation velocity on flapping wing models (Model A and Model B), it maybe prudent to develop an aerodynamic model for both models, which is the focus of this section. The backbone of the adopted aerodynamic model was proposed by Berman and Wang (2007) \cite{berman_wang_2007}, who was studying the energy-minimizing kinematics in hovering insect flight. However, this aerodynamic model only applies to FWMAV with two wings (Model A), i.e., no wing-wing interactions. Armanini et al. \cite{armanini2016quasi} extended the applicability of this model to FWMAVs with four wings (Model B). 

\subsection{Aerodynamic Model For 2 wings}\label{2wingsmodel}

\begin{figure}
		\centering
		\includegraphics[width=0.5\linewidth]{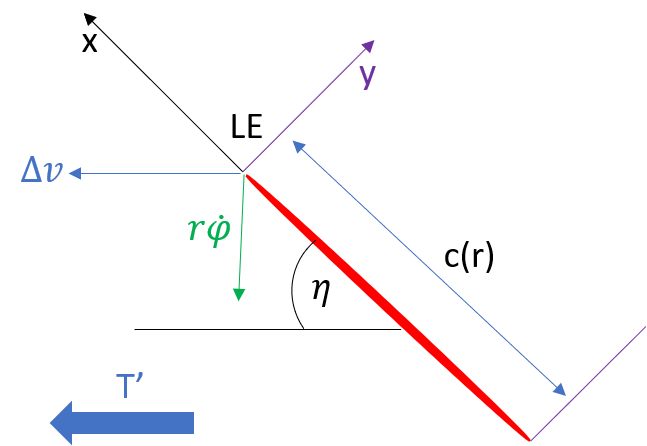}
		\caption{Cross section of the flapping wing at distance r from the body }
		\label{fig:blade}
\end{figure}
Figure \ref{fig:blade} shows a cross section (blade element) of a wing at distance $r$ from the body of the FWMAV. The red portion represents the chord of the section $c(r)$ and $(x,y)$ makes the reference frame at the leading edge (denoted as LE) of the blade element. The green and blue arrows at the leading edge denotes the direction of perturbation velocity (i.e., due to vibration) $\Delta v$ and the flapping velocity $r\Dot{\phi}$, respectively. The thrust generated per unit span is denoted as $T^{\prime}$ and can expressed by the force generated per unit span in the $x$ and $y$ directions, $F^{\prime}_{x}$, $F^{\prime}_{y}$ and the pitching angle $\eta$ 
\begin{align}\label{eqn4}
    T^{\prime}=dT/dr= F^{\prime}_{x} \operatorname{cos} \eta-F^{\prime}_{y} \operatorname{sin} \eta
\end{align}
According to Berman and Wang \cite{berman_wang_2007} the forces $F^{\prime}_{x}$ and $F^{\prime}_{y}$ can be expressed in terms of the bound vortex $\Gamma$, added masses $m_{11}$ and $m_{22}$, the velocity and acceleration components $v_{x}$, $v_{y}$, $a_{x}$ \& $a_{y}$ and the viscous forces in those directions $F^{\prime v}_{x}$ and $F^{\prime v}_{y}$ 
\begin{align}\label{Fx}
    F^{\prime}_{x}=-\rho \Gamma v_{y}-m_{11} a_{x}-F^{\prime v}_{x}
\end{align}
\begin{align}\label{Fy}
F^{\prime}_{y}=\rho \Gamma v_{x}-m_{22} a_{y}-F^{\prime v}_{y}
\end{align}
The bound vortex $\Gamma$ can be written in terms of the translational and rotational coefficients $C_{t}$ \& $C_{R}$, the total velocity $|v|=\sqrt{v^{2}_{x}+v^{2}_{y}}$, as well as the angle of attack $\alpha$ and pitching rate $\dot{\eta}$. The relation is shown in equation \ref{Gamma}. 
\begin{align}\label{Gamma}
\Gamma=-\frac{1}{2} C_{t} c(r)|v| \operatorname{sin} 2 \alpha+\frac{1}{2} C_{R} c^{2}(r) \dot{\eta}
\end{align}
The viscous force $F^{\prime v}$ are given by
\begin{align}\label{visc}
    F^{\prime v}=\frac{1}{2} \rho c(r) C_{D}|v|<v_{x}, v_{y}>
\end{align}
The added mass terms are given by

\begin{align}
    m_{11}=\frac{1}{4} \pi \rho a^{2} \; \; \; \; \; \; \& \; \; \; \; \; \; m_{22}=\frac{1}{4} \pi \rho c^{2}(r)
\end{align}
The coefficient of drag is written as
\begin{eqnarray}
C_{D}=2 C_{t} \sin ^{2} \alpha
\end{eqnarray}
The angle of attack $\alpha$ is given by
\begin{align}\label{al}
\alpha=\tan ^{-1}\left(\frac{v_{y}}{v_{x}}\right)
\end{align}
Where the velocity and the acceleration components are given as
\begin{eqnarray}
v_{x}=-r \dot{\varphi} \operatorname{Sin} \eta  \label{velo1}
\\
v_{y}=-r \dot{\varphi} \operatorname{Cos} \eta  \label{velo2}
\\
a_{x}=\dot{v}_{x}   \label{acc1}
\\
a_{y}=\dot{v}_{y}   \label{acc2}  
\end{eqnarray}
$\phi (t)$ is the flapping angle of the wing which is measured using the previously mentioned motion capture system. $\eta (r,t)$ is the pitching angle of the wing which is a function of $r$ (spanwise location) and $t$ (time). The pitching angle is modeled as a Fourier series in time since the motion is periodic, with spatially varying coefficients to account for wing flexibility. 
\begin{eqnarray}
\eta(r, t)=&& a_{0}(r)+a_{1}(r) \cos (2 \pi f t)+b_{1}(r) \sin (2 \pi f t)+a_{2}(r) \cos (4 \pi f t) +b_{2}(r) \sin (4 \pi f t)
\end{eqnarray}
Cubic polynomials are assumed for these coefficients as 
\begin{eqnarray}
a_{i}=A_{i 1} r+A_{i 2} r^{2}+A_{i 3} r^{3} \label{pitchingcoeffa}
\\
b_{i}=B_{i 1} r+B_{i 2} r^{2}+B_{i 3} r^{3}  \label{pitchingcoeffb}
\end{eqnarray}
\begin{figure}
		\centering
		\includegraphics[width=0.8\linewidth]{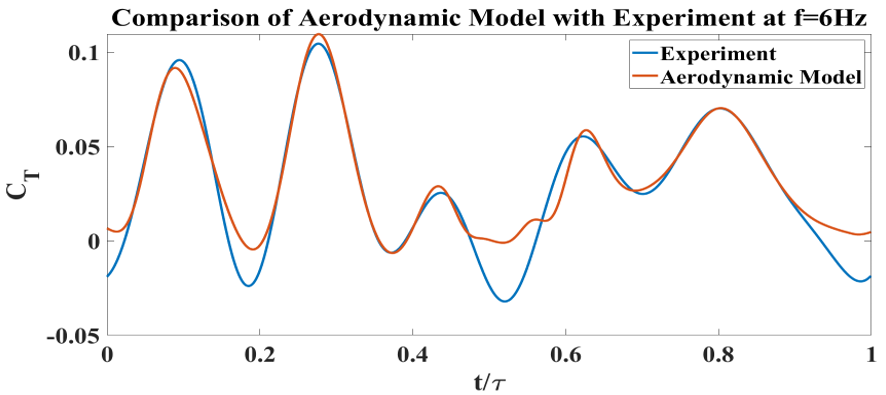}
		\caption{Comparison of aerodynamic model results with experiment at 6 Hz for 2wings}
		\label{fig:2wingresult}
\end{figure}
After obtaining thrust per unit span $T'$ from equation \eqref{eqn4}, it is then integrated over the span of the wing to determine total thrust generated by the FWMAV and then normalized to obtain the coefficient of thrust according to equation \eqref{th_2wings}.

Upon defining the complete structure of the aerodynamic model, some unknown parameters must be specified. These include  $A_{01},  A_{02}, A_{03}, A_{11}, A_{12}, A_{13}, A_{21}, A_{22}, A_{23}, B_{11}, B_{12}, B_{13}, B_{21}, B_{22}, B_{23}, C_{t} \& C_{R}$. They are determined by formulating an optimization problem to minimize the error between the theoretical prediction and the experimental measurements of thrust time-variation over the cycle using the same kinematics. The results from this optimization problem are shown in Figure \ref{fig:2wingresult}, which compares the optimized aerodynamic model with experimental measurements (the load cell test data, shown in Figure \ref{fig:lc}, are used in this case). As it can be seen, the resulting coefficient of thrust from the model and from experimental setup have a close match over the majority of the cycle. The measured perturbation velocity $\Delta v$ can be applied to the model to study the effect of induced vibrations during the flapping cycle. To apply this perturbation, we need to modify the components $v_{x}$ and $v_{y}$ of each airfoil section to account for the contribution of $\Delta v$. The modified velocities are written as, 
\begin{align}\label{pvx}
v_{x}=\Delta v \operatorname{Cos} \eta-r \dot{\varphi} \operatorname{Sin} \eta
\end{align}
\begin{align}\label{pvy}
v_{y}=-\Delta v \operatorname{Sin} \eta-r \dot{\varphi} \operatorname{Cos} \eta
\end{align}

\subsection{Aerodynamic Model of 4 wings}\label{4wingsmodel}
In case of the four-wings, the aerodynamic model is almost the same as that of the two-wings with a few extensions to capture the effect of wing-wing interaction. During the 'peel' motion, a suction is created between the wings, which sucks air from the ambient towards it and strengthens the leading edge vortex. Therefore, the generated circulation is modified in a way that empirically captures the change in the strength of the leading edge vortex.
\begin{equation}\label{gammaint}
\begin{aligned}
\Gamma &=-\frac{1}{2} C_{t} c(r)|v| \operatorname{Sin} 2 \alpha+\frac{1}{2} C_{F} c^{2}(r) \dot{\eta}_{\text {fling }}, \quad \dot{\eta}_{\text {fling }}>0 \\
\Gamma &=-\frac{1}{2} C_{t} c(r)|v| \operatorname{Sin} 2 \alpha+\frac{1}{2} C_{R} c^{2}(r) \dot{\eta}
\end{aligned}
\end{equation}
 
Also the added mass term $m_{22}$ is modified to take into account the 'peeled away' portion of the chord \cite{armanini2016quasi}. 

\begin{equation}\label{m2mod}
m_{22}=\frac{1}{4} \pi \rho c_{e f f}^{2}(r,t)
\end{equation}

\begin{figure}
		\centering
		\includegraphics[width=0.8\linewidth]{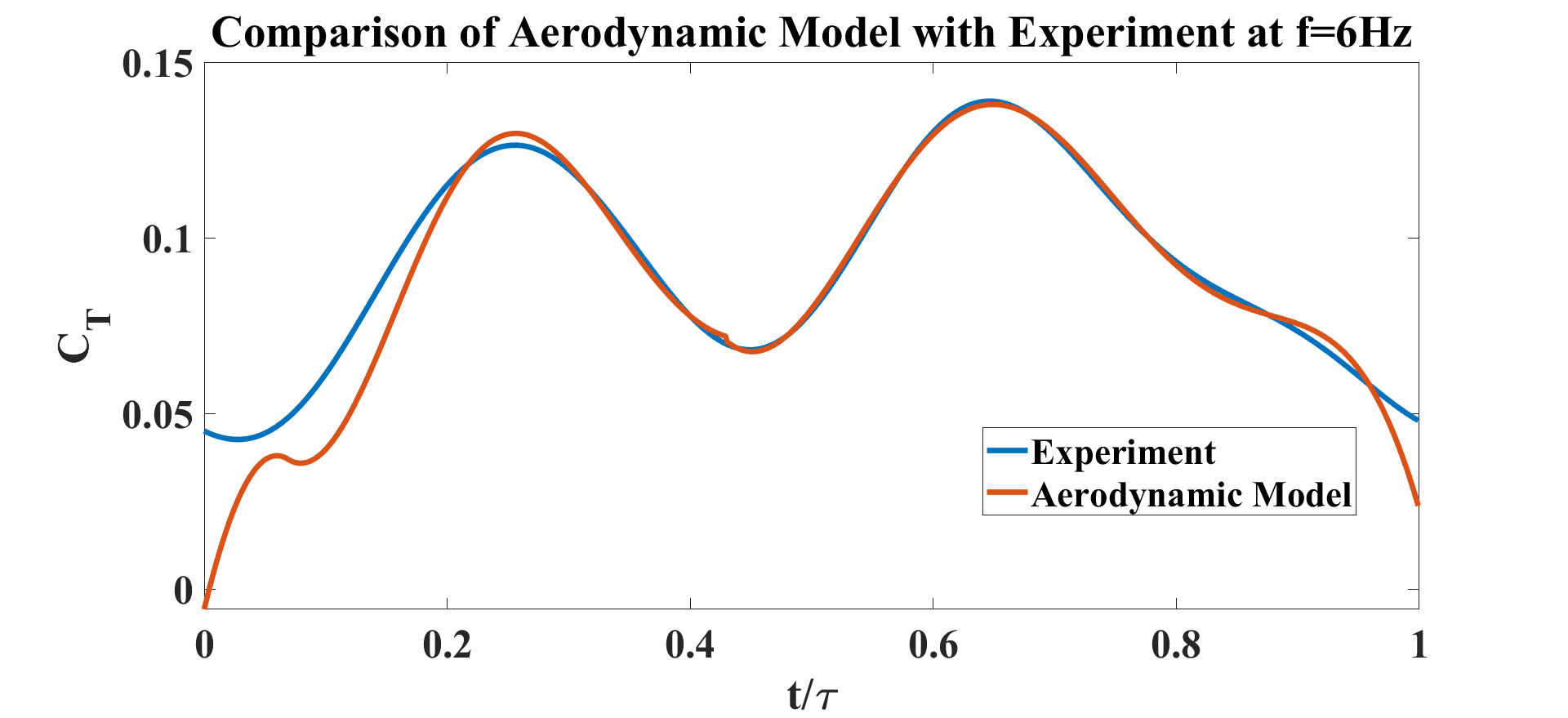}
		\caption{Comparison of aerodynamic model results with experiment at 6 Hz for 4wings}
		\label{fig:4wingresult}
\end{figure}

The result of the match of the optimization formulation is shown in Figure \ref{fig:4wingresult}. Similar to the previous case, the measured $\Delta v$ can be applied to this model to analyze the effect of vibration during the flapping cycle. 

\section{Results and Discussion}

In this section we have analyzed the flow field and the vibration measurement to explain the observation presented in Figure \ref{fig:observation}. 25\% and 35\% of the spanwise locations on the wing are chosen for flow visualization of model A. The sectional flapping velocity at the above mentioned locations are presented simultaneously with the perturbation velocity at a given point during the flapping cycle. For model B 15\% spanwise position was chosen and similar flow field images and velocities are investigated. Reference speed $V_{ref}=2\pi f R \Phi$is used to non-dimensionalize both the sectional velocity $r\Dot{\phi}$ and the induced velocity $\Delta v$.

\subsection{Effects of Self-Induced Vibrations on the Two-Wings Model (Model A)}

\begin{figure}
		\centering
		\includegraphics[width=1\linewidth]{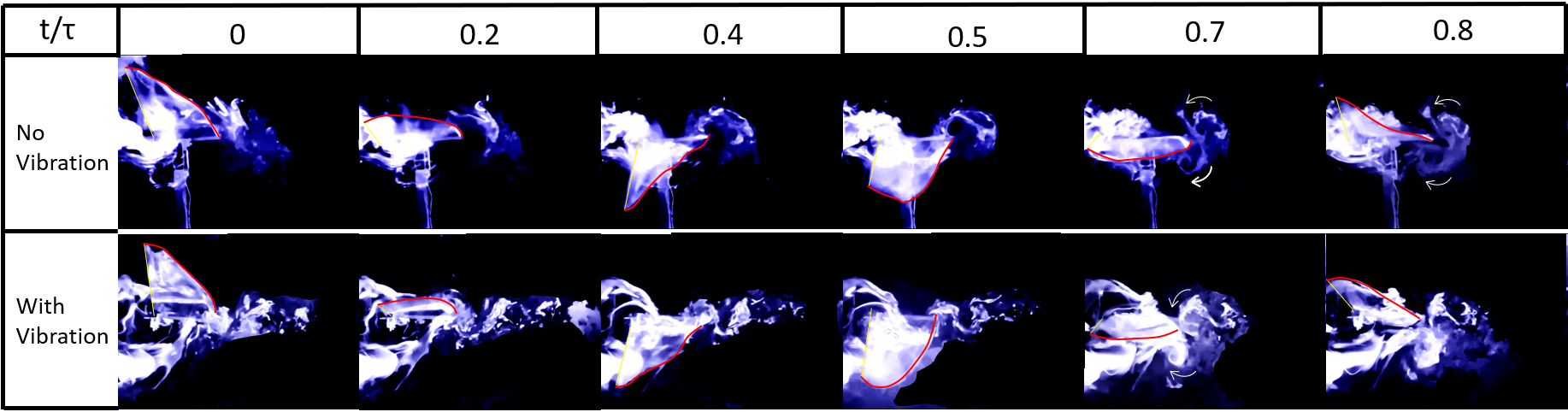}
		\caption{Flow visualization images from oscillatory test (with vibration) \& fixed test (no vibration) at 25\% spanwise location for Model A}
		\label{fig:Aflow25}
\end{figure}

\begin{figure}
		\centering
		\includegraphics[width=0.8\linewidth]{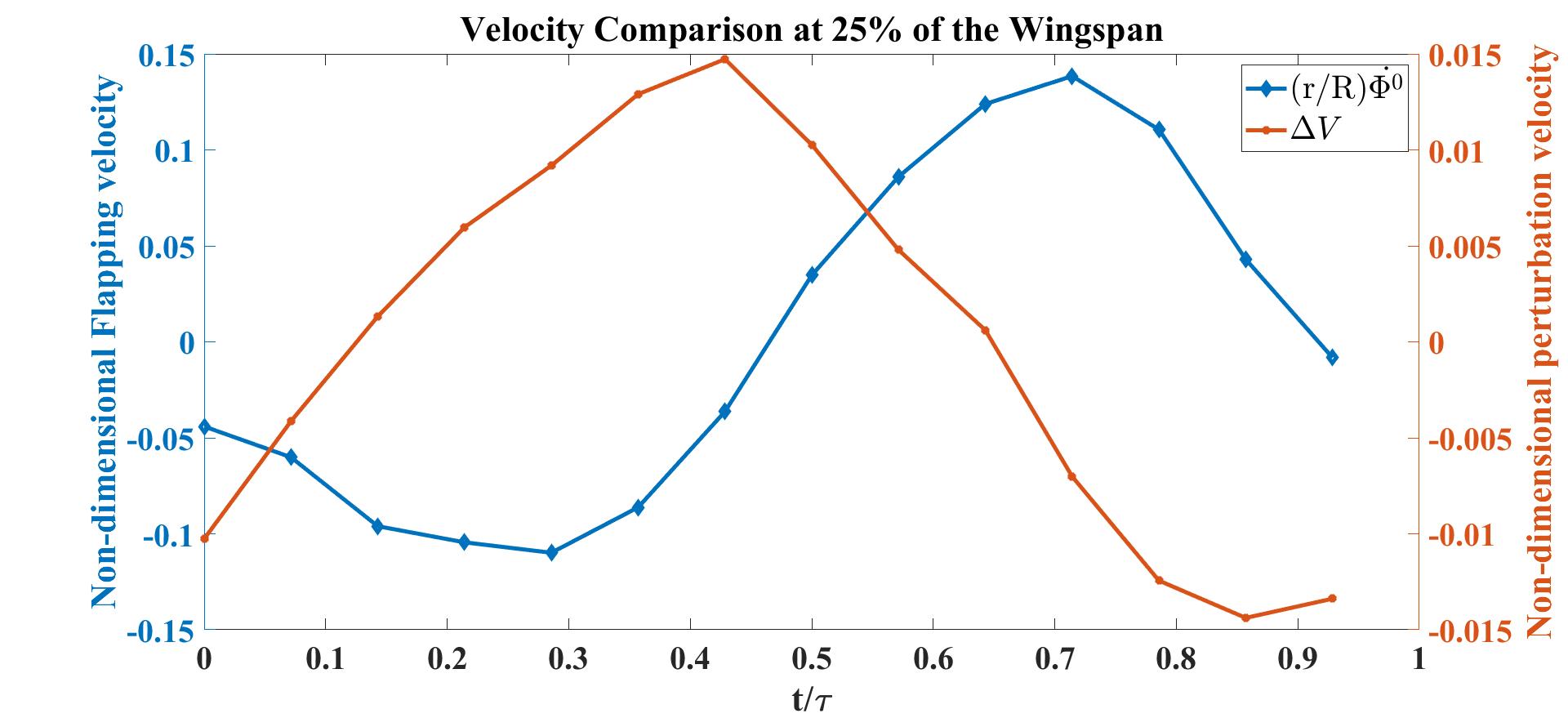}
		\caption{Normalized vibration and flapping velocity at 25\% spanwise location for Model A}
		\label{fig:Aper25}
\end{figure}

The images presented in Figure \ref{fig:Aflow25} primarily compares the flow fields produced by the flapping of Model A, at 25\% spanwise location: $1^{st}$ row is without any perturbation and the $2^{nd}$ row is with the self induced vibrations, shown in Figure \ref{fig:Aper25}. Both cases are compared at $(f=6Hz)$ flapping frequency. The row presents flow fields at a given time during the flapping period. The time parameter is denoted by $t$ where as $\tau$ time period of flapping. In the figure red shows trailing edge outline and yellow is used for leading edge. Figure \ref{fig:Aper25} shows the comparison between non-dimensional flapping velocity and normalized perturbation velocity at 25\% wingspan. The flapping cycle begins at $t/\tau=0$, when the wing starts its down-stroke near the maximum angle of flapping. The wing finishes down-stroke $t/\tau=0.5$ and ensues into upstroke. 

\begin{figure}
		\centering
		\includegraphics[width=1\linewidth]{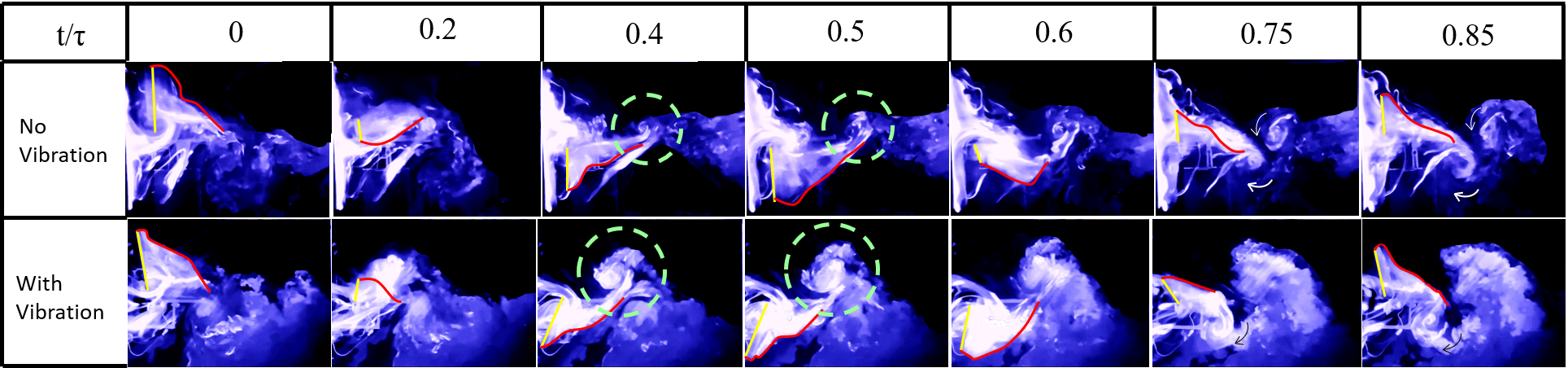}
		\caption{Flow visualization images from oscillatory test (with vibration) \& fixed test (no vibration) at 35\% spanwise location for Model A}
		\label{fig:Aflow35}
\end{figure}

\begin{figure}
		\centering
		\includegraphics[width=0.8\linewidth]{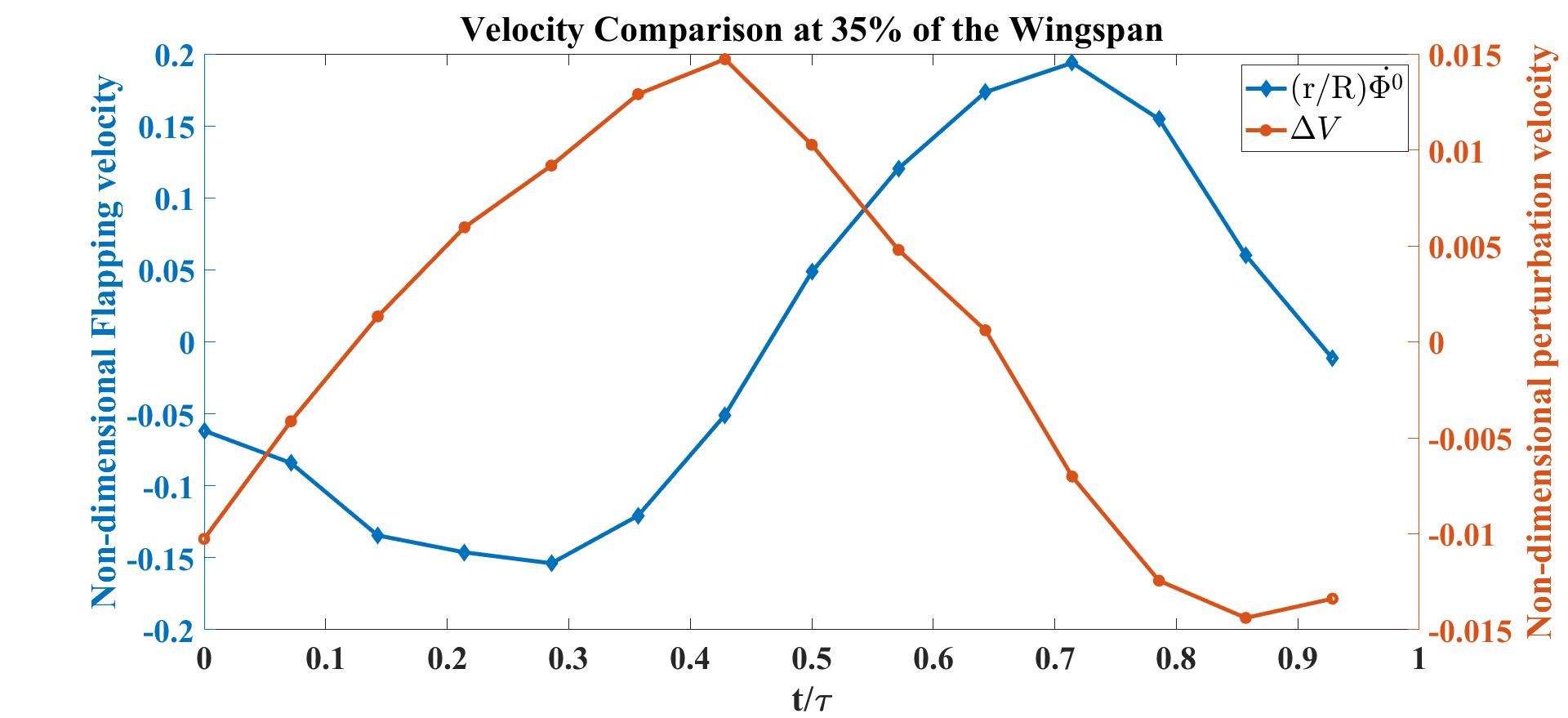}
		\caption{Normalized vibration and flapping velocity at 35\% spanwise location for Model A}
		\label{fig:Aper35}
\end{figure}

Figure \ref{fig:Aflow25} shows a couple of vortices with opposite rotations near the TE for all the cases at the instant $t/\tau=0.7$. The white arrows denotes the direction of rotation of the vortices. The similar pair of vortices with opposite rotations can be seen in the no vibration case at $t/\tau=0.8$ but they disappear in the case with vibration at similar time instant. This pair of vortices at the trailing edge indicates presence of a jet, which favors thrust generation. Figure \ref{fig:Aper25} shows a negative perturbation velocity at $t/\tau = 0.7 \; \& \; 0.8$, which implies a motion of the FWMAV model towards the jet due to the self-induced vibration. That is, the whole body is moving towards the counter-rotating vortices, which ebbs the jet effect and consequentially decreases the thrust in the oscillatory case. Similar physics can be observed at 35\% of the wing-span, which is shown in Figure \ref{fig:Aflow35}. Also the normalized velocity comparison at 35\% wing-span is shown in Figure \ref{fig:Aper35}.

\begin{figure}
		\centering
		\includegraphics[width=0.8\linewidth]{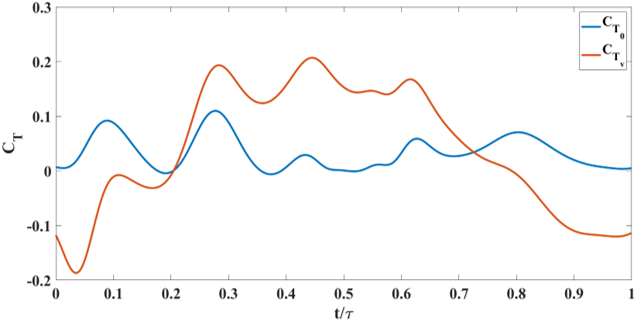}
		\caption{Thrust coefficient comparison for Model A with and without self-induced body vibration}
		\label{fig:comp2wings}
\end{figure}

Figure \ref{fig:comp2wings} shows the coefficient of thrust during a flapping cycle for model A, with and without self-induced vibration. $C_{T_{0}}$ denotes the thrust coefficient based on the the aerodynamic model discussed in  section \ref{2wingsmodel} \cite{berman_wang_2007} without any perturbation, while $C_{T_{v}}$ is the coefficient of thrust when the measured vibration-induced perturbation is applied to the aerodynamic model. Figure \ref{fig:comp2wings} shows that approximately in between $t/\tau=0.2 \; \& \; 0.7$, $C_{T_{v}}$ is greater than $C_{T_{0}}$. The reason can be seen in the flow visualization. Presented in Figure \ref{fig:Aflow35}; that at $t/\tau=0.4 \; \& \; 0.5$,  in the no vibration case, there is a single trailing edge vortex (TEV) attached to the trailing edge. By contrast in the oscillatory case TEV is detached from the trailing edge. The TEVs are shown in green dashed circles. Model A generates thrust using a conventional unsteady lifting mechanism: A TEV is shed whenever there is a change in the wing motion, which changes the wing bound circulation as well as aerodynamic forces, because of the conservation of circulation. This mechanism has a transient response (Wagner's effect \cite{wagner1925uber}). Due to this transient, the closer the TEV to the trailing edge, the smaller its strength compared to the steady value. Figure \ref{fig:Aflow35} shows that at $t/\tau=0.4 \; \& \; 0.5$, $\Delta V$ is positive, which means that the whole body, due to the perturbation is moving to the left, leaving the TEV detached from the trailing edge. This results in an increase in $C_{T_{v}}$ over a specific duration. But after $t/\tau=0.7$, $\Delta V$ becomes negative and the whole body moves into the jet and loses thrust as discussed previously. This is reflected in Figure \ref{fig:comp2wings} that beyond $t/\tau >0.7$, $C_{T_{v}}$ becomes way less than $C_{T_{0}}$. At the end, the average $C_{T_{v}}$ is less than the average $C_{T_{0}}$ over the flapping cycle. % ($\overline{{{C_{T_{0}}}}}=0.0393$ \& $\overline{{{C_{T_{v}}}}}=0.0316$).

%\subsection{Discussion on Model B}

\subsection{Effects of Self-Induced Vibrations on the Four-Wings Model (Model B)}

\begin{figure}
		\centering
		\includegraphics[width=1\linewidth]{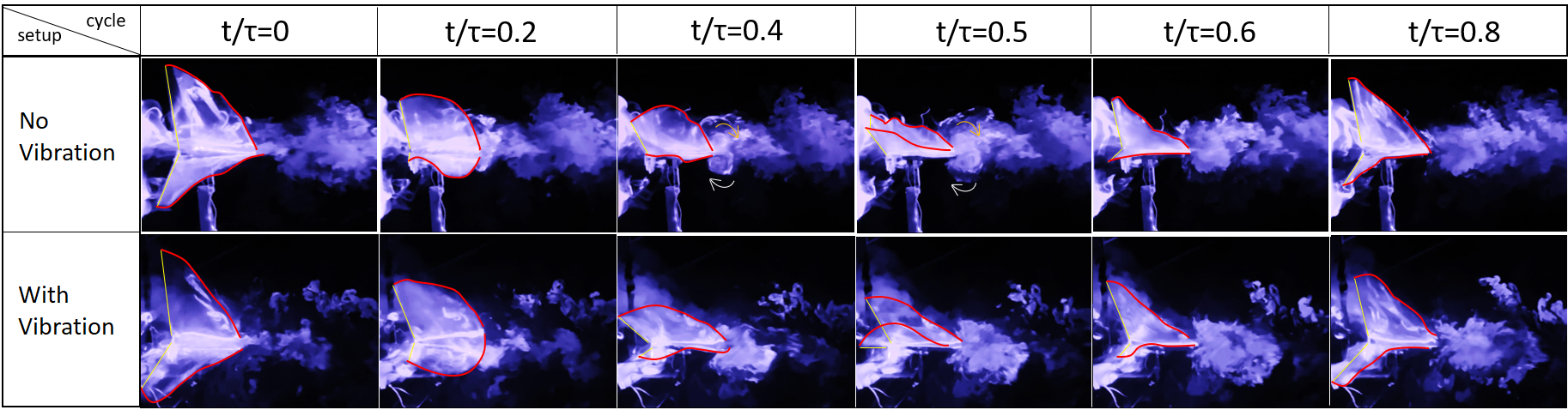}
		\caption{Flow visualization images from oscillatory test (with vibration) \& fixed test (no vibration) at 15\% spanwise location for Model B}
		\label{fig:Bflow15}
\end{figure}

\begin{figure}
		\centering
		\includegraphics[width=0.8\linewidth]{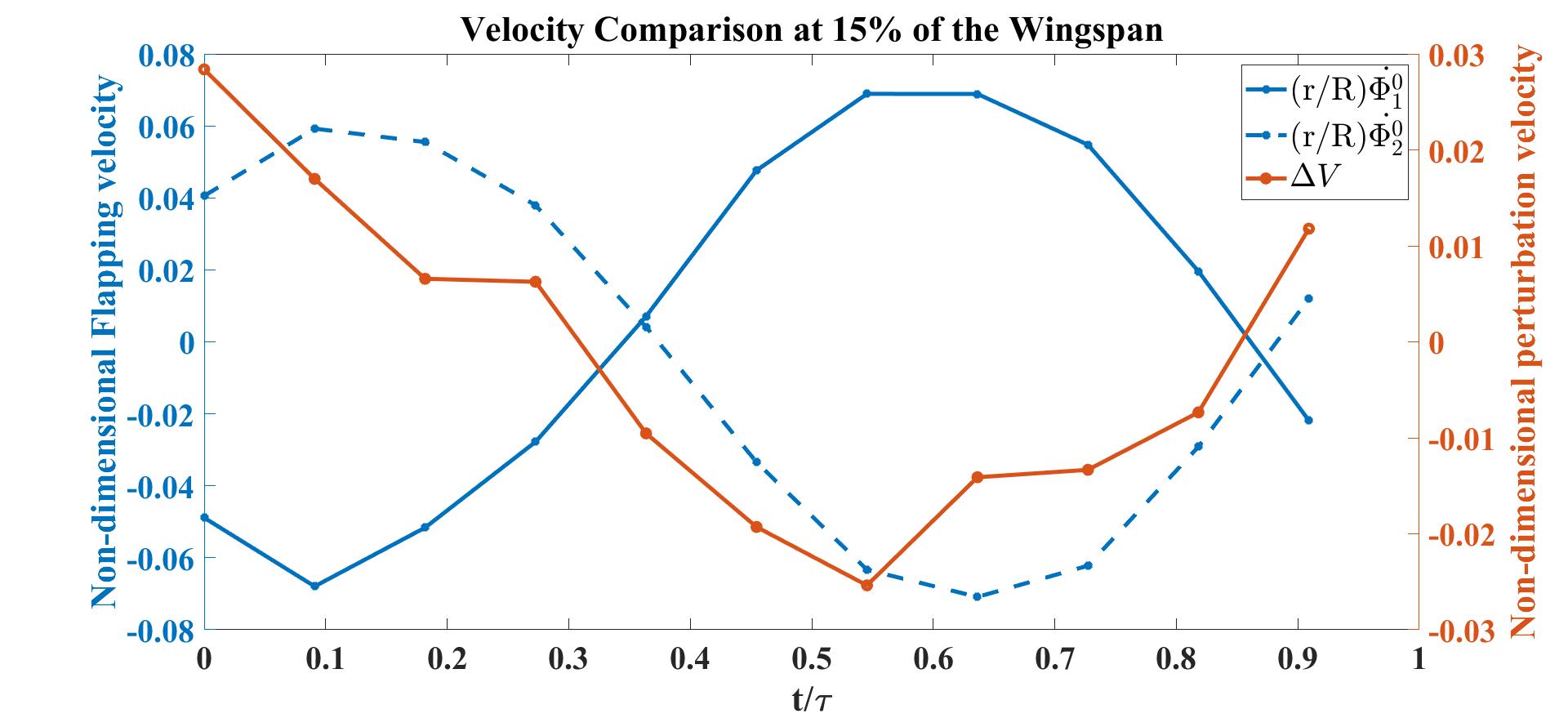}
		\caption{Normalized vibration and flapping velocity at 15\% spanwise location for Model B}
		\label{fig:Bper15}
\end{figure}

\begin{figure}
		\centering
		\includegraphics[width=0.8\linewidth]{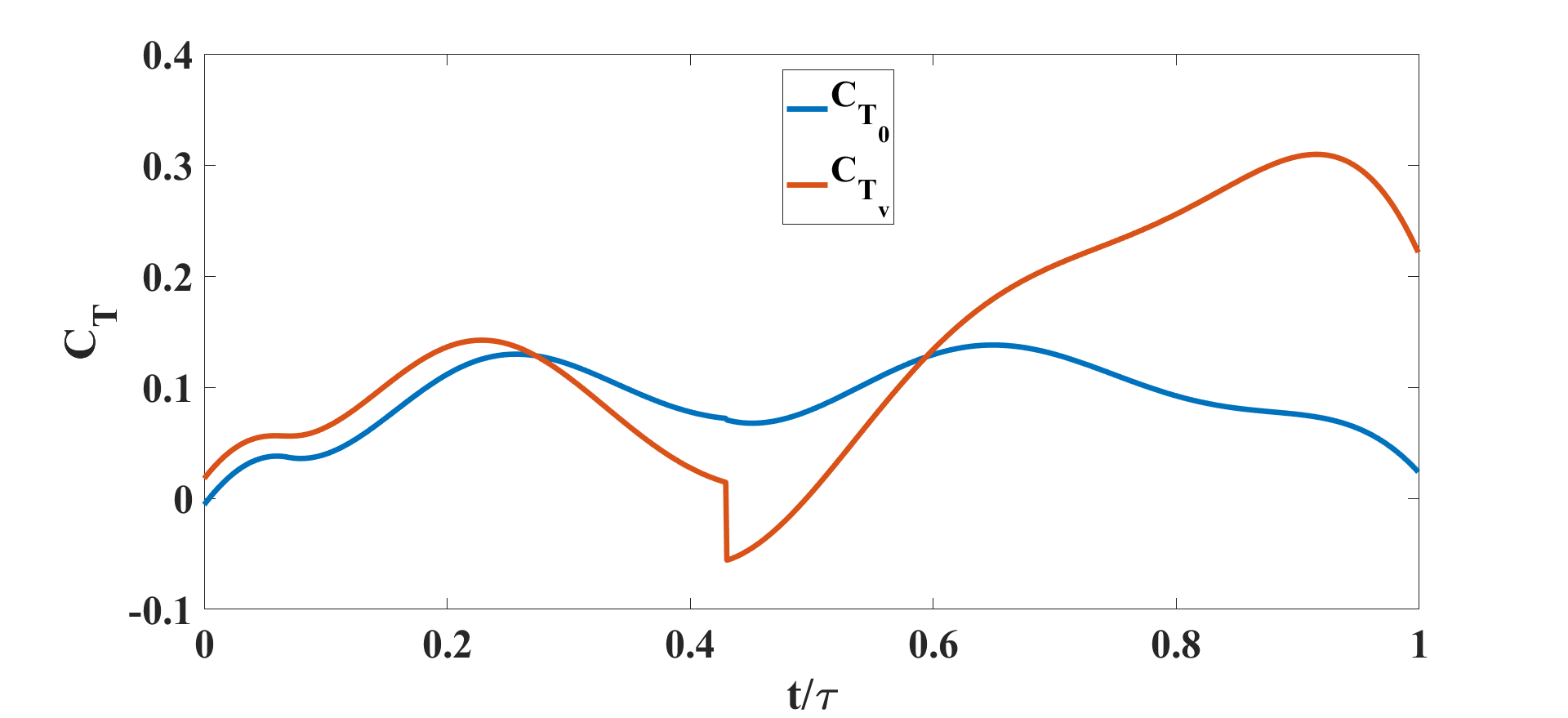}
		\caption{Thrust coefficient comparison for Model B with and without self-induced body vibration}
		\label{fig:comp4wings}
\end{figure}

This sub-section is dedicated to analyze the effect of self-induced vibration in the flow field of Model B at 15\% spanwise location. As mentioned by Balta et al. \cite{MiquelDipan2021}, the four-wings mechanism generates thrust through the clap and peel mechanism. During the peel motion, it creates a suction between the wings which intakes a significant amount of air. During the clap motion, it pushes the air downstream creating a 'jet burst'. Figure \ref{fig:Bflow15} shows that for the no vibration case, there are two counter rotating vortices at $t/\tau=0.4 \hspace{1mm} \& \hspace{1mm} 0.5$ which are indicative of the 'jet burst'. However, the figure does not show similar vortices in the oscillatory case at the same instant. Figure \ref{fig:Bper15} shows that the perturbation velocity at $t/\tau=0.4 \hspace{1mm} \& \hspace{1mm} 0.5$ is negative. This implies that the FWMAV is moving to the right during this time. This motion of the FWMAV towards the jet decreases the thrust, which can be seen in Figure \ref{fig:comp4wings}. In this figure, $C_{T_{v}}$ is the thrust coefficient resulting from the aerodynamic model presented in section \ref{4wingsmodel} with the perturbation and $C_{T_{0}}$ is the coefficient without the perturbation. We can see that $C_{T_{v}}$ is less than $C_{T_{0}}$ at $t/\tau=0.4 \hspace{1mm} \& \hspace{1mm} 0.5$. In contrast, Figure \ref{fig:Bper15} shows that $\Delta V$ is increasing during the ensuing period which means the flapping robot is moving away from the jet. The jet burst and the self-induced perturbation takes the vortices away from the trailing edge. Thus it enhances the clapping effect; and the thrust increases significantly, as shown in Figure \ref{fig:comp4wings} after $t/\tau =0.6$, compared to the case with no vibration. This enhanced clapping effect dominates the average $C_{T_{v}}$ over the average $C_{T_{0}}$.

%\begin{figure}
%		\centering
%		\includegraphics[width=0.5\linewidth, %height=0.5\linewidth]{Picture1.png}
%		\caption{Schematic of Thrust Enhancement for Model B due to self-induced body vibration}
%		\label{fig:explanationB}
%\end{figure}

In the table below the experimental and aerodynamic modeling values of $\frac{\overline{C_{T_{v}}}-\overline{C_{T_{0}}}}{\overline{C_{T_{0}}}}$ are presented in percentage format for flapping frequency 6Hz.

\begin{center}
\begin{table}
    
\begin{tabular}{||c c c||} 
\hline
    & 2 wings & 4 wings \\ [0.5ex] 
 \hline\hline
 Experimental & -59.1\% & 64.4\% \\ 
 \hline
 Aerodynamic Modeling & -24.4\% & 55.6\% \\ [1ex]
   
 \hline

\end{tabular}
\centering
\caption{Percentage of average thrust coefficient change from ideal hovering to the case with perturbation, using aerodynamic modeling and experimental measurement}
\end{table}
\end{center}

\section{Conclusion}
The current study compares two cases of hovering flights. The first case is of an ideal hovering; there is no room for perturbation. The measurement in this case is carried out using loadcell setup, which is also called the fixed test. The other case of hovering provides with the concomitant self-induced vibration due to the oscillatory nature of the thrust force. A pendulum setup, which is also called the oscillatory test, is used to measure the average force in this case. This force measurement includes the effect of the self-induced vibration. These cases are studied using two different flapping wing robots: a two-wings robot or model A and a four-wings robot or model B. When these models are tested in the above mentioned setups, it is observed that the fixed test measures more thrust than the oscillatory test for model A. The opposite behaviour is observed for model B as shown in Figure \ref{fig:observation}. The Model B exploits 'clap-and-peel' for generating thrust, where Model A uses the conventional flapping mechanism for the same. Due to the difference in the thrust generation mechanisms, the effect of the perturbation also differs between the two models. 

Two well known aerodynamic models (\cite{armanini2016quasi} \cite{berman_wang_2007}) for thrust generation are used to match the loadcell data by optimizing some unknown parameters. The perturbation velocity, measured using the motion capture system, is applied in the model to study the effect of the induced vibration during the flapping cycle. It is obvious that the perturbation has some effect on the flow field as well. To investigate how much impact the vibration has on the flow field, flow visualization technique is used to look into it. 

Flow visualization revealed some interesting vortex interactions. In the case with no vibration, model A enjoys a certain jet effect near its trailing edge. The perturbation wanes its effect by moving the whole flapping robot into the jet. This decreases the overall thrust for model A in the oscillatory test. For model B the self induced vibration enhances the thrust by moving the flapping robot away from the jet. This phenomenon enhances the clapping effect and consequently increases overall thrust in the oscillatory test. The vortex interactions show how the self induced vibration has an adverse effect on the thrust generation for a two-wings flapping robot. In contrast, the vibration has enhancing effect on a four-wings flapping robot.

\section*{Acknowledgment}
We want to thank Ming Shao for his valuable suggestions in the completion of the project. And we would also like to thank the National Science Foundation and the Air Force Office of Scientiﬁc Research for funding this research work.

%\section*{Declarations}

%\subsection{Ethical Approval}
%Not applicable

%\subsection{Competing interests}
%Not applicable

%\subsection{Authors' contributions}
%Dipan Deb and Haithem E Taha contributed to conceptualizing the study. Dipan Deb and Kevin Huang built the experimental setup and acquired data. Aakash Verma helped in writing the post-processing code. Moatasem Fouda helped in writing the LabView code. Dipan Deb wrote the manuscript. Dipan Deb and Haithem E Taha reviewed the manuscript.

%\subsection{Funding}
%We would like to thank the National Science Foundation (Grant number CMMI-1846308) for funding this research work.

%\subsection{Availability of data and materials}
%The data and other materials will be available upon reasonable request. 

\bibliography{sample}

\end{document}